\documentclass[aps, pra, longbibliography,superscriptaddress]{revtex4-1}
\usepackage[english]{babel}
\normalsize 

\usepackage{amsmath,amsfonts}
\usepackage{rotating}
\usepackage{leftidx}
\usepackage{here}
\usepackage{pbox}
\usepackage[braket,qm]{qcircuit}
\usepackage[nolist,nohyperlinks]{acronym}
\usepackage{physics}

\usepackage{float}
\usepackage[dvipsnames]{xcolor}
\usepackage{graphicx}
\usepackage{caption}
\usepackage{subcaption}
\usepackage{tikz}
\usetikzlibrary{decorations.pathreplacing,calligraphy}
\usepackage{natbib}
\usepackage{dcolumn} 
\usepackage{multirow}
\usepackage{booktabs}

\newcommand{\fett}[1]{\mbox{\boldmath$#1$}}
\newcommand{\rr}{ \fett r}
\newcommand{\xx}{ \fett x}
\newcommand{\RR}{ \fett R}
\newcommand{\acr}[1]{\hat {\textmd a}_{#1}^{\dagger}}
\newcommand{\aan}[1]{\hat {\textmd a}_{#1}}

\renewcommand{\vec}[1]{\boldsymbol{#1}}
\newcommand{\tR}[1]{\textcolor{red}{ #1}}
\newcommand{\tB}[1]{\textcolor{blue}{ #1}}

\newcommand{\vtheta}{\vec{\theta}}
\newcommand{\vect}{\vec{t}}

\newcommand{\creation}{\hat{a}^{\dagger}} 
\newcommand{\anni}{\hat{a}} 
\newcommand{\tht}[1]{\tilde \theta_{#1}}

\begin{document}

\title{A Quantum Computing Implementation of Nuclear-Electronic Orbital (NEO) Theory: Towards an Exact pre-Born-Oppenheimer Formulation of Molecular Quantum Systems}

\author{Arseny Kovyrshin} 
\email{arseny.kovyrshin@astrazeneca.com}
\affiliation{ Data Science and Modelling, Pharmaceutical Sciences, R\&D, AstraZeneca Gothenburg, Pepparedsleden 1, Molndal SE-431 83, Sweden}
\author{M{\aa}rten Skogh} 
\email{marten.skogh@astrazeneca.com}
\affiliation{ Data Science and Modelling, Pharmaceutical Sciences, R\&D, AstraZeneca Gothenburg, Pepparedsleden 1, Molndal SE-431 83, Sweden}
\author{Anders Broo}
\email{anders.broo@astrazeneca.com} 
\affiliation{ Data Science and Modelling, Pharmaceutical Sciences, R\&D, AstraZeneca Gothenburg, Pepparedsleden 1, Molndal SE-431 83, Sweden}

\author{Stefano Mensa}
\email{stefano.mensa@stfc.ac.uk}
\author{Emre Sahin}
\email{emre.sahin@stfc.ac.uk}
\affiliation{The Hartree Centre, STFC, Sci-Tech Daresbury, Warrington, WA4 4AD, United Kingdom}
\author{Jason Crain}
\email{jason.crain@ibm.com}
\affiliation{IBM Research Europe, Hartree Centre STFC Laboratory, Sci-Tech Daresbury, Warrington WA4 4AD, United Kingdom}
\affiliation{Department of Biochemistry, University of Oxford, Oxford, OX1 3QU, UK}
\author{Ivano Tavernelli}
\email{ita@zurich.ibm.com}
\affiliation{IBM Quantum, IBM Research Europe – Zurich, S\"aumerstrasse 4, 8803 R\"uschlikon, Switzerland}

\begin{abstract}

 Nuclear quantum phenomena beyond the Born–Oppenheimer approximation are known to play an important role in a growing number of chemical and biological processes. While there exists no unique consensus on a rigorous and efficient implementation of coupled electron-nuclear quantum dynamics, it is recognised that these problems scale exponentially with system size on classical processors and therefore may benefit from quantum computing implementations. Here, we introduce a methodology for the efficient quantum treatment of the electron-nuclear problem on near-term quantum computers, based upon the \acl{neo} (NEO) approach. We  generalize the electronic two-qubit tapering scheme to include nuclei by exploiting symmetries inherent in the 
 NEO framework; thereby reducing the hamiltonian dimension, number of qubits, gates, and measurements needed for calculations. We also develop parameter transfer and initialisation techniques, which improve convergence behavior relative to conventional initialisation. These techniques are applied to \acs{h2} and malonaldehyde for which results agree with \acl{neofci} and \acl{neocasci} benchmarks for ground state energy to within $10^{-6}$ Ha and entanglement entropy to within $10^{-4}$. These implementations therefore significantly reduce resource requirements for full quantum simulations of molecules on near-term quantum devices while maintaining high accuracy. 
\end{abstract}
\maketitle

\begin{acronym}
\acro{vrte}[VRTE]{Variational Real-Time Evolution}
\acro{vite}[VITE]{Variational Imaginary-Time Evolution}
\acro{nisq}[NISQ]{Noisy Intermediate-Scale Quantum}
\acro{vha}[VHA]{Variational Hamiltonian Ansatz}
\acro{bo}[BO]{Born--Oppenheimer}
\acro{pbo}[pBO]{pre-Born--Oppenheimer}
\acro{nomo}[NO+MO]{Nuclear Orbital plus Moloecular Orbital}
\acro{neo}[NEO]{Nuclear-Electronic Orbital}
\acro{neohf}[NEOHF]{Nuclear Electronic Orbitals Hartree--Fock}
\acro{hf}[HF]{Hartree--Fock}
\acro{sto}[STO-3G]{single-$\zeta$ (minimal) basis set contracted from 3 Gaussian primitives}
\acro{sto6}[STO-6G]{single-$\zeta$ (minimal) basis set contracted from 6 Gaussian primitives}
\acro{631}[6-31G]{split-valence double-$\zeta$ Gaussian basis set in 6-31 contraction scheme}
\acro{dzsnb}[DZSNB]{split-valence double-$\zeta$ nuclear basis set composed of 2 uncontracted Cartesian $S$ functions}
\acro{vqe}[VQE]{Variational Quantum Eigensolver}
\acro{avqe}[AdaptVQE]{Adaptive Variational Quantum Eigensolver}
\acro{neocasci}[NEOCASCI]{Nuclear-Electronic Orbital Complete Active Space Configuration Interaction}
\acro{casci}[CASCI]{Complete Active Space Configuration Interaction}
\acro{neofci}[NEOFCI]{Nuclear-Electronic Orbital Full Configuration Interaction}
\acro{fci}[FCI]{Full Configuration Interaction}
\acro{ucc}[UCC]{Unitary Coupled Cluster}
\acro{neoucc}[NEOUCC]{Nuclear-Electronic Orbitals Unitary Coupled Cluster}
\acro{uccsd}[UCCSD]{Unitary Coupled Cluster Singles and Doubles}
\acro{neouccsd}[NEOUCCSD]{Nuclear-Electronic Orbitals Unitary Coupled Cluster Singles and Doubles}
\acro{lih}[Li-H]{lithium hydride}
\acro{h2}[H$_2$]{hydrogen molecule}
\acro{c3h5o2}[C$_3$H$_5$O$_2$]{protonated enol malonaldehyde form}
\acro{qpe}[QPE]{Quantum Phase Estimation}
\acro{cobyla}[COBYLA]{Constrained Optimization By Linear Approximation}
\acro{cg}[CG]{Conjugate Gradient}
\acro{slsqp}[SLSQP]{Sequential Least SQuares Programming}
\acro{mp2}[MP2]{the second-order M\o{}ller--Plesset}
\acro{ts}[TS]{transition state}
\end{acronym}

\newpage

\section{Introduction}

Chemical simulations at the atomic scale have advanced to the point where they are now capable of describing a wide range of phenomena including those occurring in complex systems and biological contexts.  
Much of this progress is due to advances in the power and efficiency of methods that can capture the quantum mechanical nature of electrons.  
Common to nearly all practical implementations of these techniques are two foundational simplifying assumptions:  (a) that the atomic nuclei behave as classical particles and (b) that the \ac{bo} approximation holds wherein the electrons are assumed to move sufficiently fast that they respond adiabatically to the nuclear motion~\cite{markland2018nuclear}.
 
It is however now widely acknowledged that, in systems containing light atoms, nuclear quantum effects are not negligible  and  can make appreciable contributions to processes including proton delocalization and tunneling, occurring \textit{e.g.}, in biological systems~\cite{pereyaslavets2018importance} including enzymatic catalysis~\cite{heyes2009nuclear}, in tautomeric transitions, and in determining the relative stability of crystal polymorphs~\cite{rossi2016anharmonic}.  
The observation of an isotope-dependence in reaction rate~\cite{vardi2015nuclear} is a further particularly striking signature of nuclear quantum effects. 
Even in a familiar substance such as liquid and solid water, the presence of light atoms involved in hydrogen bonding~\cite{raugei2003nuclear} suggests that nuclear quantum effects can give important contributors to its static and dynamical properties~\cite{kim2017temperature,pamuk2012anomalous}.
Finally, phenomena such as charge transfer reactions occurring for instance in photo-chemistry (\textit{e.g.}, in the light harvesting complexes) may involve situations where the electronic potential energy surfaces (PES) become fully degenerate. At these so-called conical intersections, nuclear motion couples to more than a single PES and the adiabatic approximation breaks down opening new radiationless decay pathways.  These examples provide a strong motivation for the extension of the current theoretical frameworks beyond the \ac{bo} and classical nuclei approximations. 

Currently, several methodologies have been developed that give a full quantum description of chemical systems. They are often referred to as the \ac{neo} approach~\cite{webb2002}, \ac{pbo} quantum theory~\cite{simm2013}, or \ac{nomo} theory~\cite{naka2002}. 
In these formulations, nuclear tunneling and isotope effects arise naturally. 
Additionally, phenomena involving breakdown of the \ac{bo} approximation are also captured. 
However, the computational cost of exact nuclear-electronic structure methods implemented on classical computers increases exponentially with the system size. 
In contrast, quantum processors have the potential to reduce this cost scaling to polynomial-time --- underpinning the concept of quantum advantage~\cite{lee2022evidence,kjaergaard2020superconducting,huang2020superconducting}. 

Quantum computing is emerging as a new computational paradigm for the efficient solution of quantum mechanical problems, which can have a tremendous impact in different domains such as 
many-body physics~\cite{Somma2002,wecker2015,bauer2016hybrid,smith_simulating_2019,cade2020, 
chiesa_quantum_2019,RevTacchino2020, Suchsland_2021, miessen2021quantum, PhysRevResearch.4.043038, PhysRevResearch.4.043011}, 
high energy physics~\cite{martinez_real_time_2016,klco_quantum-classical_2018,roggero_dynamic_2019,mathis_toward_2020,mazzola2021gauge,wu2021,crippa2022,PhysRevResearch.3.033221,schuhmacher2023unravelling,wozniak2023quantum}, 
quantum chemistry~\cite{OMalley2016, kandala2017hardware, Barkoutsos2018, cao2019quantum,McArdle2020,sokolov2020, ollitrault2019, sokolov_microcanonical_2021,kiss2022quantum, barkoutsos2021quantum,ollitrault2021molecular,rossmannek2021quantum}, material design~\cite{Panos_Alchemical_2021}, 
biology and medicine~\cite{robert2021resource,mensa2022quantum, baiardi2022quantum,maniscalco2022quantum}. 
The first quantum algorithm for treatment of both nuclei and electrons in a fully quantum simulation setting was proposed by Kassal \textit{et al.}~\cite{kass2008} and applied to the calculation of the quantum dynamics for the hydrogen molecule.  
Later, Veis \textit{et al.}~\cite{veis2016b} adopted the \ac{nomo} approach for constructing molecular Hamiltonians and proposed a \acs{qpe} algorithm for computing the corresponding ground state wave function. 
Finally,  Ollitrault \textit{et al.}~\cite{olli2020} presented a quantum algorithm for the simulation of nonadiabatic electron-nuclear dynamics including excited states. 

The \acs{neo} approach is particularly versatile and stands as a compromise between the \ac{bo} approximation and methodologies such as \ac{pbo}. 
In the \acs{neo} approach, light mass nuclei are treated using orbital techniques in the same way as are the electrons. 
Furthermore, the BO separation between electronic and nuclear wave functions is --- in general --- not necessary in this case. 
On the contrary, the heavy nuclei (usually everything heavier than hydrogen) are described by classical point charges, determining the geometry of the molecular scaffold.  
Pavošević and Hammes-Schiffer~\cite{pavo2021} recently proposed a quantum computing implementation of the \ac{neo} approach employing the \ac{vqe} algorithm for the optimization of the hydrogen molecule and positronium hydride ground state wave functions, where only one proton was treated at quantum level.

In this study, we investigate the potential of the \ac{neo} approach for chemical systems and focus on the construction of the corresponding qubit Hamiltonian and wave function ans\"atze with the aim of accessing larger system sizes than have been previously investigated.
Specifically, we explore and implement improved wave function parameterizations, operational mappings of second quantized molecular Hamiltonians to the qubit space, and efficient parameter initialization schemes for several electron-nuclear wave function ans{\"a}tze. 
Particularly important are the devise of qubit tapering schemes such as two-qubit reduction, the exploitation of molecular point group symmetries, and the application of projectors to efficiently sample the relevant sectors of the molecular Fock space. 

After a detailed description of the \acs{neo} Hamiltonian and of the corresponding wave function in the quantum computing setting, we tested our algorithm on the evaluation of the ground state energies of the hydrogen and malonaldehyde molecules using exact --- noise-free --- state vector \ac{vqe}  simulators. 
In the case of malonaldehyde, our approach was able to predict --- despite the quite small size of the nuclear basis set and the imposed rigidity of the molecular scaffold --- proton transfer barriers in reasonable good agreement with the available references. 
Exploiting the full quantum nature of the molecular wave function, we also proposed methods for the evaluation of  additional interesting quantum features, such as the entanglement entropy of the subsystems (electronic and nuclear) from the full \ac{neo} wave function.

\section{Theory}

\subsection{\acl{neo} Approach}
Relative to the standard \ac{bo} approximation, with point charges representing nuclei, the Hamiltonian in the \ac{neo} approach includes additional kinetic and potential energy operators for selected light nuclei. 
This approach incorporates the quantum effects of light nuclei and additionally 
introduces a separation between heavy and light nuclear motion. 
We denote the collective spatial coordinates of all quantum objects as $\rr$, and additionally introduce superscripts where required to distinguish between nuclei, $\rr^n$, and electrons, $\rr^e$. We refer to both quantum nuclear and electronic coordinates combined as $\rr = \{\rr^e, \rr^n\}$.
The coordinates for classical nuclei will be denoted as $\RR$. 
Quantities associated with electrons will carry lower case indices $i,j$, while upper case indices, $I, J$, are used for nuclear quantities, regardless of whether they are classical or not. 
The total wave function, $\Psi(\rr;\RR)$, will still retain a parametric dependence on classical nuclear coordinates, $\RR$. 
Thus the Schr{\"o}dinger equation reads as follows (in atomic units)
\begin{align}\label{eq:fqSchroed}
\left[
-\sum_{i}^{N^e}\frac{1}{2}\nabla^2_i
-\sum_{I}^{N^n}\frac{1}{2M_I}\nabla^2_I
-\sum_{i,I}^{N^e,N^n}\frac{Z_{I}}{\left|\rr_i-\rr_I\right|}
+\sum_{i < j}^{N^e}\frac{1}{\left|\rr_{i}-\rr_{j}\right|}
-\sum_{i,A}^{N^e,N^c}\frac{Z_{A}}{\left|\rr_i-\RR_A\right|} \right. \nonumber\\
\left.
+\sum_{I, A}^{N^n,N^c}\frac{Z_{I}Z_{A}}{\left| \rr_{I}-\RR_{A}\right|}
+\sum_{A < B}^{N^c}\frac{Z_{A}Z_{B}}{\left| \RR_{A}-\RR_{B}\right|}
\right]
\Psi(\rr;\RR)=
E\Psi(\rr;\RR),
\end{align}
where $Z_I$ are nuclear charges, $M_I$ nuclear masses, $N^e$, $N^n$, and $N^c$ are the total number of electrons, quantum nuclei, and classical nuclei, respectively.

The \acs{neo} approach solves Eq.\ (\ref{eq:fqSchroed}) using both electronic,
\begin{equation}\label{eq:soe}
\phi_i(\xx_i) \equiv \phi_i(\rr_i,s_i),
\end{equation}
and nuclear, 
\begin{equation}\label{eq:son}
\phi_I(\xx_I) \equiv \phi_I(\rr_I,s_I),
\end{equation}
spin orbitals constructed separately from nuclear and electronic basis sets, originally in Gaussian basis sets~\cite{webb2002}. Note that the spin coordinates $s_i$ in Eqs.\,(\ref{eq:soe},\ref{eq:son}) are combined together with $\rr_i$ into a new variable $\xx_i$ for brevity. 
Based on these spin orbitals one can form antisymmetric products, Slater determinants, 
for electrons
\begin{equation}\label{eq:elProd}
\Phi_{\mu}(\xx^{e})= \hat S^- \prod_{i}\phi_i(\xx^e_i),
\end{equation}
and symmetric or antisymmetric  spin orbital products for bosonic or fermionic nuclei, correspondingly,
\begin{equation}\label{eq:nucProd}
\Phi_{\nu}(\xx^{n})= \hat S^{\pm}\prod_{I}\phi_I(\xx^n_I),
\end{equation}
where $\hat S^{+/-}$ are symmetrizer/antisymmetrizer operators. These constitute the multi-particle basis set for the \ac{neofci} approach~\cite{webb2002}, which expresses the nuclear-electronic wave function as follows
\begin{align}\label{eq:neofci}
\Psi(\xx;\RR) = \sum_{\mu \nu}^{C^e,C^n}C_{\mu \nu} \Phi_{\mu}(\xx^{e}) \Phi_{\nu}(\xx^{n}) \, ,
\end{align}
where $C^e$ and $C^n$ are the total number of electronic and nuclear states, Eq.~\eqref{eq:elProd} and \eqref{eq:nucProd}. 
We emphasize that in Eq.~\eqref{eq:neofci} only one type of nuclei is considered. 
In the most general case, each type of nuclei will have its own set of spin orbitals and corresponding symmetric or antisymmetric product, which in turn enters the sum in Eq.~\eqref{eq:neofci}. 
As in this work, we will only consider protons and electrons as quantum particles, only masses $M_I=M= 1874 \, m_e$ and charges $Z_I=1\,e$ will be considered in the quantum part of the Hamiltonian. 
Within such a framework, one gets the following second quantization representation for the Hamiltonian~\cite{webb2002}
\begin{multline}
\hat H =
-\sum_{ij} \left[ \int \phi^*_{i}(\xx) \frac{1}{2}\nabla^2 \phi_{j}(\xx) d\xx \right] \acr{i} \aan{j}
-\sum_{IJ} \left[\int \phi^*_{I}(\xx) \frac{1}{2M}\nabla^2 \phi_{J}(\xx) d\xx \right] \acr{I} \aan{J} \\[2ex]
+\frac{1}{2}\sum_{ijkl} \left[\int \phi^*_i(\xx_1) \phi^*_{k}(\xx_2) \frac{1}{|\rr_1-\rr_2|}\phi_{l} (\xx_2)
\phi_{j} (\xx_1) d\xx_1 d\xx_2 \right] \acr{i} \acr{k} \aan{l} \aan{j} \\[2ex]
+ \frac{1}{2}\sum_{IJKL} \left[ \int \psi^*_I(\xx_1) \psi^*_{K}(\xx_2) \frac{1}{|\rr_1-\rr_2|}\psi_{L} (\xx_2)
\psi_{J} (\xx_1) d\xx_1 d\xx_2 \right] \acr{I} \acr{K} \aan{L} \aan{J} \\[2ex]
-\sum_{ijKL} \left[\int \phi^*_i(\xx_1) \psi^*_K(\xx_2) \frac{1}{|\rr_1-\rr_2|}\psi_L (\xx_2) \phi_j (\xx_1) d\xx_1 d\xx_2 \right]
\acr{i} \acr{K} \aan{L} \aan{j} \\
+\sum_{IJ,A} \left[\int \phi^*_{I}(\xx) \frac{Z_A}{|\rr-\RR_A|} \phi_{J}(\xx) d\xx \right] \acr{I} 
\aan{J}
-\sum_{ij,A} \left[\int \phi^*_{i}(\xx) \frac{Z_A}{|\rr-\RR_A|} \phi_{j}(\xx) d\xx \right] \acr{i} \aan{j}\\
+\frac{1}{2}\sum_{AB} \frac{Z_AZ_B}{|\RR_A-\RR_B|}.
\end{multline}
Note that in addition to anti-commutation relations between indistinguishable fermions
\begin{align}
\left[ \acr{i}, \aan{j} \right]^+ &= \delta_{ij}\\
\left[ \acr{i}, \acr{j} \right]^+ &= \left[ \aan{i}, \aan{j} \right]^+   = 0\\
\left[ \acr{I}, \aan{J} \right]^+ &= \delta_{IJ}\\
\left[ \acr{I}, \acr{J} \right]^+ &= \left[ \aan{I}, \aan{J} \right]^+   = 0
\end{align}
distinguishable fermions do commute (between protons and electrons), implying
\begin{align}
\left[\acr{I}, \aan{j}\right]   &= 0 \label{eq:dcom3}\\ 
\left[\acr{I}, \acr{j}\right]   &= \left[\aan{I}, \aan{j}\right]     = 0. \label{eq:dcom4}
\end{align}
If bosonic nuclei are considered then the following commutation 
relations hold for indistinguishable particles
\begin{align}
\left[\acr{I}, \aan{J}\right] &= \delta_{ij},\\
\left[\acr{I}, \acr{J}\right] &= \left[\aan{I}, \aan{J}\right]  = 0
\end{align}
and the same relation as above, Eqs.\,(\ref{eq:dcom3},\ref{eq:dcom4}), between the distinguishable particles.

\subsection{Quantum Computing Ansatz}\label{sec:ansaetze}

 The \acs{vqe} is  a quantum variational algorithm that uses a classical optimization routine to find the parameters of a quantum state that minimizes the expectation value of a given molecular Hamiltonian \cite{peru2014}.
 Some of the key advantages of the VQE algorithm include its ability to be implemented on near-term quantum computers  which have a limited number of qubits and are prone to errors, and its potential to scale to larger systems, hence becoming a tool for practical applications in the near future \cite{moll2018,kand2017,seeley2012bravyi,mcardle2020quantum}. 

Mathematically, the VQE optimization algorithm can be expressed as follows. Let $\hat{H}$ be the Hamiltonian of the system and let $\ket{\psi(\boldsymbol{\theta})}$ be the trial state, which is a function of parameters $\boldsymbol{\theta}$. The goal is to find the values of $\boldsymbol{\theta}$ that minimize the expectation value of the Hamiltonian, given by
\begin{equation}
E(\boldsymbol{\theta}) = \bra{\psi(\boldsymbol{\theta})} \hat{H} \ket{\psi(\boldsymbol{\theta})}.
\end{equation}

To find the optimal values of $\boldsymbol{\theta}$, we can use a classical optimization algorithm to iteratively update the values of $\boldsymbol{\theta}$. Then, the quantum part of
the algorithm is used to apply the quantum circuit to the quantum state of the system and measure the energy of the resulting state until the minimum value of $E(\boldsymbol{\theta})$ is found.

For the \ac{vqe} optimization of the system wave function within a quantum computing setting, it is desirable to design a flexible, expressive and well-behaved ansatz, which can deliver good accuracy with the fewest number of parameters. 
While there are a vast number of different ans{\"a}tze available, in this work we focus on the \acl{ucc} and hardware-efficient parameterizations. 
The main advantage of the latter arises from the shallow circuit depth required for its implementation --- an important prerequisite for near-term (non-fault tolerant) hardware~\cite{preskill2018quantum}. 
Unfortunately, for chemical systems, there is no systematic, practical approach for the initialization and optimization of the variational parameters when using hardware-efficient ans{\"a}tze. 
Moreover, these often show limited expressivity while the optimization process suffers from the presence of barren plateaus~\cite{mcclean2018barren}. Here we will explore several variants of the TwoLocal hardware-efficient ansatz~\cite{qiskit} for the \ac{neo} wave function; details of which will be discussed in Section~\ref{sec:partran}.

In contrast to hardware-efficient ans{\"a}tze, the \ac{ucc} ansatz has a very well-tailored parameter space and naturally approximates wave functions for chemical systems. 
This well-established quantum chemistry ansatz was first introduced for \ac{vqe} by Peruzzo \textit{et al.}~\cite{peru2014}. 
Although the resulting quantum circuits are in general deeper than the hardware-efficient ones, 
the optimization of the \acs{ucc} circuit is less prone to barren plateaus, and lack of particle, total spin, and spin projection conservation. 
This becomes especially important when one uses multiple centers for protons and electrons, leading hardware-efficient ansatz optimizations to the wrong results with an incorrect number of electrons and nuclei.

While a hardware-efficient ansatz does not require any special adaptation for the \ac{neo} wave 
function, we must introduce tailored modifications to the \ac{ucc} ansatz to extend it from the purely electronic to the \ac{neo} case. Here we refer to the modified forms as \ac{neo}\ac{ucc}. 
The parameterized unitary operators $\hat{U}_{CC}(\vect)$, 
\begin{equation}
    \hat{U}_{CC}(\vect) = e^{\hat{T}(\vect)-\hat{T}^\dagger(\vect)}
\end{equation}
are at the core of the \ac{ucc} ansatz. 
Acting upon some initial state $\ket{\psi_{init}}$ (typically the Hartree-Fock solution) it produces a trial wave function of the form
\begin{equation}
\ket{\psi(\vect)} = \hat{U}_{CC}(\vect)\ket{\psi_{init}} .
\end{equation}
In general,  the operator $\hat{T}$ is given as a sum of cluster operators of increasing order, \textit{i.e.}, the number of particles being acted on, $o$,
\begin{equation}\label{eq:ucc sum}
    \hat{T} = \sum_o \hat{T}^{(o)} = \hat{T}^{(1)} + \hat{T}^{(2)} + \ldots
\end{equation}
In other words, $o=1$ corresponds to one-particle excitations, $o=2$ to two-particle excitations, and so on. In second quantization these are expressed as
\begin{align}
    \hat{T}^{(1)} &= \sum_{pq} t^{p}_{q} \, \creation_p \anni_q \\
    \hat{T}^{(2)} &= \sum_{pqrs} t^{pq}_{rs} \, \creation_p \anni_r  \creation_q \anni_s \, .
\end{align}
Here $p,q,r,s$ index all spin orbitals, both protonic and electronic, and $t$ represents coefficients to be optimized.
Labeling the initially occupied spin orbitals with $i,j,k,l$ and the initially unoccupied spin orbitals with $a,b,c,d$ (analogous for protons using upper case: $I,J,K,L$ and $A,B,C,D$) we can limit the cluster operators to a subset that conforms to chosen occupations of the initial state 
\begin{equation}
    t^{ab\ldots AB\ldots}_{ij\ldots IJ\ldots} \prod_{\gamma, \lambda} \creation_\gamma \anni_\lambda \prod_{\Gamma, \Lambda} \creation_\Gamma \anni_\Lambda,
\end{equation}
where $\gamma \subseteq \{a,b,\ldots\}$, $\lambda \subseteq \{i,j,\ldots\}$, $\Gamma \subseteq \{A,B,\ldots\}$, $ \Lambda \subseteq \{I,J,\ldots\}$.
That is, annihilation (creation) operators will only act on occupied (unoccupied) spin orbitals, indexed as $a,b,c,d,\ldots$ ($i,j,k,l,\ldots$) for electronic spin orbitals and as $A,B,C,D,\ldots$ ($I,J,K,L,\ldots$) for protonic spin orbitals. 
Since we now obtain a mix of electron and proton operators we introduce a dual superscript $\hat{T}^{(e,p)}$ where the first index, $e$, is the order of electronic excitation in each operator, and $p$ is similarly the order of the protonic excitations, that is
\begin{equation}\label{eq:ucc ep sum}
    \hat{T} = \sum_{e,p} \hat{T}^{(e,p)},
\end{equation}
where $\hat{T}^{(e,p)}$ is \textit{e.g.}
\begin{align}
    \hat{T}^{(1,0)} &= \sum_{ia} t^{a}_{i} \, \creation_a \anni_i\\ 
    \hat{T}^{(0,1)} &= \sum_{IA} t^{A}_{I} \, \creation_A \anni_I \\
    \hat{T}^{(2,0)} &= \sum_{ijab} t^{ab}_{ij} \, \creation_a\anni_i \creation_b\anni_j\\
    \hat{T}^{(0,2)} &= \sum_{IJAB} t^{AB}_{IJ} \, \creation_A \anni_I \creation_B \anni_J\\
    \hat{T}^{(1,1)} &= \sum_{iJaB} t^{aB}_{iJ} \, \creation_a \anni_i \creation_B\anni_J
\end{align}

The sum in Eq.~\eqref{eq:ucc ep sum} is normally truncated to some fixed order of $e$ and $p$. Electronic structure \acs{vqe} calculations are often limited to one- and two-particle cluster operators, while higher orders are ignored. This approximation is commonly referred to as \ac{uccsd}. We will extend this notation to indicate when different orders are used for electrons, protons and mixed terms, respectively. For example, an ansatz using singles and doubles for electrons and nuclei, while using a mixed double consisting of single electronic and single nuclear excitations, a mixed triple consisting of double electronic, and single nuclei operators will be indicated as
\begin{equation}
    \text{UCCSDT$^{(2,1)}$}\Rightarrow \overbrace{\hat{T}^{(1,0)} + \hat{T}^{(0,1)}}^{\text{S}}  + 
    \underbrace{\hat{T}^{(2,0)} + \hat{T}^{(1,1)} + \hat{T}^{(0,2)}}_{\text{D}} + \hat{T}^{(2,1)} \, .
\end{equation}

Note that the lack of superscript implies use of all $o$-particle operators, \textit{i.e.}, $\text{S}\Rightarrow\hat{T}^{(1,0)} + \hat{T}^{(0,1)}$, $\text{D}\Rightarrow\hat{T}^{(2,0)} + \hat{T}^{(1,1)} + \hat{T}^{(0,2)}$, $\text{T}\Rightarrow\hat{T}^{(3,0)} + \hat{T}^{(2,1)} + \hat{T}^{(1,2)} + \hat{T}^{(0,3)}$, \textit{etc.} 
For systems with fewer electrons or protons than the excitation order, $n_e < e + p$ or $n_n < e + p$, only mixed operators are considered. For example, in the case of the hydrogen molecule, the possible triple and quadruple excitation operators are $\text{T} \Rightarrow \hat{T}^{(2,1)} + \hat{T}^{(1,2)}$ and $\text{Q} \Rightarrow \hat{T}^{(2,2)}$, respectively.

\subsection{Ansatz Initialization}\label{sec:ansatzinit}

The initial guess for parameters in any variational quantum algorithm is in many cases important to the performance of the calculation and can be crucial. Having a proper initial guess can reduce the number of evaluations needed to reach convergence, while also allowing convergence to the correct minimum (or maximum). As such, it proves wise to choose initial values for the ansatz parameters with care. For hardware-efficient ansätze this is especially vital, where otherwise initial parameters often are chosen randomly, making convergence to a global minimum difficult.

\begin{figure}[htb]
    \centering
    \begin{tikzpicture}[scale=1.5]

    \def\qey{0.5}  
    \def\qpy{-0.5} 

    \def\innerboxyoffset{0.4}
    \def\outerboxyoffset{0.6}
    
    \draw (0,\qey) node (q0) [left] {$\ket{q_e}$};
    \draw (0.3,\qey+0.07) node {$/^{n_e}$};
    \draw (0,\qpy) node (q2) [left] {$\ket{q_p}$};
    \draw (0.3,\qpy+0.07) node {$/^{n_p}$};
    
    \draw[thick]  (1,\qey+\outerboxyoffset)   rectangle (5.6,\qpy-\outerboxyoffset); 
    \draw[dashed] (1.2,\qey+\innerboxyoffset) rectangle (3.3,\qey-\innerboxyoffset);
    \draw[dashed] (1.2,\qpy+\innerboxyoffset) rectangle (3.3,\qpy-\innerboxyoffset);
    \draw[dashed] (3.5,\qey+\innerboxyoffset) rectangle (5.4,\qpy-\innerboxyoffset);
    
    \draw[] (q0) -- (1, \qey);
    \draw[] (q2) -- (1,\qpy);
    
    \draw[] (5.6, \qey) -- +(1,0);
    \draw[] (5.6,\qpy) -- +(1,0);
    
    \draw (2.3,\qey) node [] {$U_{T^{(e,0)}}(\vtheta^{e})$}  ;
    \draw (2.3,\qpy) node [] {$U_{T^{(0,n)}}(\vtheta^{n})$}  ;
    \draw (4.5,0.0)  node [] {$U_{T^{(e,n)}}(\vtheta^{e,n})$};
    
    \end{tikzpicture}
    \caption{Schematic circuit used for the initialization of the electronic parameters of \ac{neo}\ac{ucc} ansatz. The conventional electronic operators can be considered as a subset of the \ac{neo} operators. $n_e$ and $n_p$ are the numbers of qubits representing electron and proton spin orbitals, respectively.}
    \label{fig:UCCAnsatzGeneral}
\end{figure}
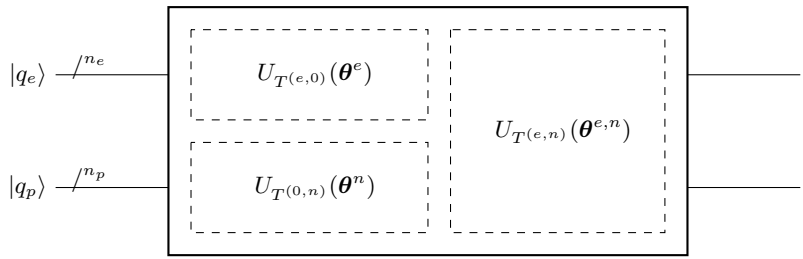

The convergence problem arises due to several factors, \textit{e.g.}, the presence of local minima in the optimization landscape, and the presence of barren plateaus~\cite{mcclean2018barren,wang2021noise}. As these problems compound with the number of variational parameters, it can be useful to solve an approximate problem in order to find a set of starting parameters closer to those of the true ground state. In the case of \ac{neo} calculations, one such approximate solution can be taken from the conventional electronic structure calculations under the \ac{bo} approximation and with nuclei represented as classical point charges. Thus we suggest initializing the parameters for the electronic part of the \ac{neo} calculation with those from the converged electronic structure calculation. For the case of \ac{neo}\ac{ucc} ansatz, the circuit operators corresponding to electronic, nuclear, and mixed excitations can be separated according to Figure~\ref{fig:UCCAnsatzGeneral}. As $U_{T^{(e,0)}}(\vtheta^{e})$ operators are the same for \ac{neo} and electronic cases one can initialize their parameters, $\vtheta^{e}$, with those from the electronic calculation and set all other parameters to zero, $\vtheta^{0,n} = \vtheta^{e,n} = \boldsymbol{0}$.

\begin{figure}[htb]
    {\centering
    \begin{tikzpicture}[scale=1]
        
        \def\neotopy{2.0}
        \def\neobottomy{-2.0}

        \def\neoleftx{1.0}
        \def\neorightx{5.0}

        \def\qey{1.0}  
        \def\qpy{-1.0} 
        
        \draw (0,\qey) node (q0) [left] {$\ket{q_e}$};
        \draw (0.3,\qey+.07) node {$/^{n_e}$};
        \draw (0,\qpy) node (q2) [left] {$\ket{q_p}$};
        \draw (0.3,\qpy+0.07) node {$/^{n_p}$};

        \draw[] (q0) -- (\neoleftx, \qey);
        \draw[] (q2) -- (\neoleftx, \qpy);
        
        \draw[] (\neorightx, \qey) -- +(1,0);
        \draw[] (\neorightx,\qpy) -- +(1,0);
        
        \draw[thick,fill=white] (\neoleftx+0.3,\neotopy-0.3) rectangle (\neorightx+0.3,\neobottomy-0.3);
        \draw[thick,fill=white] (\neoleftx+0.2,\neotopy-0.2) rectangle (\neorightx+0.2,\neobottomy-0.2);
        \draw[thick,fill=white] (\neoleftx+0.1,\neotopy-0.1) rectangle (\neorightx+0.1,\neobottomy-0.1);
        \draw[thick,fill=white] (\neoleftx+0.0,\neotopy-0.0) rectangle (\neorightx+0.0,\neobottomy-0.0);
        \draw (\neoleftx+2.0,\qpy) node [] {$U_{\text{\ac{neo}}}(\vtheta_{\text{\ac{neo}}})$};
        
        \draw[thick,dashed,fill=white] (\neoleftx+0.3,\neotopy-0.3) rectangle (\neorightx-1.0+0.3,\neotopy-2.0+0.1);
        \draw[thick,dashed,fill=white] (\neoleftx+0.2,\neotopy-0.2) rectangle (\neorightx-1.0+0.2,\neotopy-2.0+0.2);
        \draw[thick,dashed,fill=white] (\neoleftx+0.1,\neotopy-0.1) rectangle (\neorightx-1.0+0.1,\neotopy-2.0+0.3);
        \draw (\neoleftx+1.5,\qey) node [] {$U_{\text{El.}}(\vtheta_{\text{El.}})$};

         \draw [
            ultra thick,
            pen colour={Cerulean},
            decorate, 
            decoration = {calligraphic brace, 
            raise=5pt,
            amplitude=5pt},rotate around={45:(\neorightx-1.0+0.3,\neotopy-0.5)}
        ] (\neorightx-1.0+0.3,\neotopy+0.1) -- (\neorightx-1.0+0.3,\neotopy-0.5);
        \draw[] (\neorightx-1.0+0.15,\neotopy) node [above right] {$d_{\text{El.}}$};
        
        \draw [
            ultra thick,
            pen colour={orange},
            decorate, 
            decoration = {calligraphic brace,
            raise=5pt,
            amplitude=5pt},rotate around={45:(\neorightx+0.3,\neotopy-0.5)}
        ] (\neorightx+0.3,\neotopy+0.2) -- (\neorightx+0.3,\neotopy-0.5);
        \draw[] (\neorightx+0.1,\neotopy-0.1) node [above right] {$d_{\text{\ac{neo}}}$};
        
    \end{tikzpicture}}
    \caption{Schematic circuit used for the initialization of the electronic parameters in the \ac{neo} hardware-efficient ansatz. 
    The electronic and \ac{neo} hardware-efficient circuits have $d_{\text{El.}}$ and $d_{\text{\ac{neo}}}$ layers, respectively. 
    $n_e$ and $n_p$ are the numbers of qubits representing electron and proton spin orbitals.
    }
    \label{fig:TwoLocalGeneral}
\end{figure}
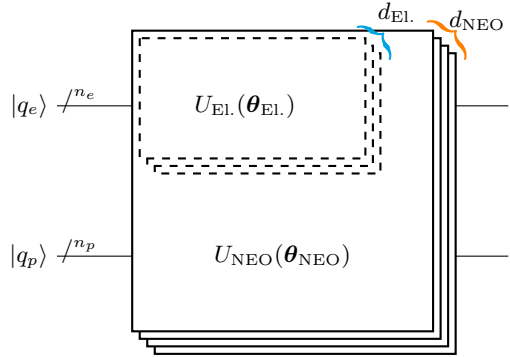

For a hardware-efficient ansatz, the parameter transfer becomes slightly more complicated as it is difficult to assign given parameters to specific nuclear or electronic parts. Therefore, special care should be taken such that the parameter transfer only affects qubits associated with electronic orbitals. A generalized construction of the TwoLocal ansatz can be created from sets of parameterized single qubit rotational gates separated by entangling gates applied between qubits. Both the electronic and \ac{neo} problems can be represented in this manner, see Figure~\ref{fig:TwoLocalGeneral}. In general, the number of layers needed to accurately describe the electronic problem, $d_{\text{El.}}$, is smaller or equal to $d_{\text{\ac{neo}}}$, the number of layers needed to describe the \ac{neo} problem. With this in mind and a proper selection of entangling scheme, one can find a transfer protocol between electronic and \ac{neo} parameters. In Section~\ref{sec:partran} we present in detail the two variants of parameter initialization employing CZ entangling layers. CZ entangling layers are beneficial in that they allow for the limited influence of electronic parameters on nuclear qubits if the nuclear parameters are unset (equal to zero). 

The suggested initialization schemes assumes that the electronic parameters could also be relaxed during the wave function optimization. It would be especially important for cases where multiple \ac{bo} potential energy surfaces intersect. Also, one should assure that the presence of additional nuclear orbitals in the \ac{neo} representation preserves the point group symmetry of the system. Nonetheless, even if the point group symmetry is the same, the corresponding $\mathbb{Z}_2$-symmetry reduction might introduce a difference in the sign pattern between the off-diagonal elements of the \ac{neo} and electronic qubit 
Hamiltonians. This stems from the difference in sign between eigenvalues of $\mathbb{Z}_2$-symmetry for the \acs{neo} and electronic states and must be taken into account during the parameter transfer. We describe  the procedure in detail in Section~\ref{sec:partran}.

\subsection{Entanglement Entropy Measurement}\label{sec:qi}

The measurement of the entanglement entropy (or von Neumann entropy) 
can provide additional insights into the correlation between the two particle subsystems, namely electrons and nuclei. 
This is particularly interesting 
when studying systems composed of distinguishable particles and will provide a direct measurement of the entanglement between the two subsystems. 
Here we will introduce the subsystem entropy in the \ac{neo} framework as  
\begin{equation}\label{eq:vonNeuman}
s_i=-{\rm Tr}( \hat \rho^{i} \ln \hat \rho^{i}) \, ,
\end{equation}
where $i \in \{e,n\}$, $\hat \rho^{e}={\rm Tr}_{n} \hat \rho^{e,n}$ and $\hat \rho^{n}={\rm Tr}_{e} \hat \rho^{e,n}$ are the corresponding reduced density matrices, and $\hat \rho^{e,n}$ is the full density matrix.

\section{Computational Details}\label{sec:CD}
The \ac{neo} framework was implemented in a locally modified version of Qiskit~\cite{qiskit}. 
All quantum computing calculations were 
performed as state vector simulations using variations of the TwoLocal 
ansatz as well as \ac{neo}\acs{ucc} ans{\"a}tze. 
The former employed parameterized $R_y$ 
and $R_z$ single rotation gates together with CZ entangling gates. 
The various entanglement patterns used in this work  are summarized in Sections 
\ref{sec:partran} and \ref{sec:h2}. 
All the classical algorithm calculations were carried out using the GAMESS-US software package~\cite{GAMESS}. 
In order to prepare the test systems and reference results the \ac{neohf}, \ac{neofci}, and \ac{neocasci}~\cite{webb2002} calculations were performed. 
The system Hamiltonians were expanded in the orbitals obtained from the \ac{neohf} calculations,
and the corresponding \ac{neo} wave functions were optimized with the \ac{vqe} algorithm in Qiskit. 
The SciPy~\cite{gommersscipy} implementations of the \ac{cobyla}~\cite{powell1994direct}, 
\ac{cg}~\cite{hest1952,nazareth2009conjugate}, and \ac{slsqp}~\cite{kraft1994slsqp} 
methods were used for the optimization of the circuit parameters.  
The fermion-to-qubit transformation used to map the electronic and nuclear operators together with the employed qubit reduction techniques are described in Section \ref{sec:mapping}. 
The additional qubit space reduction obtained by exploiting the \ac{neo} Hamiltonian's point group symmetries are discussed in Section (\ref{sec:symred}).

\subsection{Fermionic Mapping and 4-qubit Reduction}\label{sec:mapping}
The fermionic parity mapping~\cite{seeley2012bravyi} scheme 
is used in this work for both electronic and nuclear second-quantized operators. Note that we are only treating nuclei consisting of single protons quantum mechanically, and as such, only fermionic representations are necessary. We assume that the fermionic occupation number basis vectors are written in what we will refer to as particle-spin block representation
\begin{equation}
    \ket{f} = \ket*{
    \underbrace{f^{p}_{n_\beta ^{p} + n_\alpha^p },\ldots,
    f^{p}_{1+n_\alpha^p}}_{\beta \text{-spin prot.}},
     \overbrace{f^{p}_{n_\alpha^{p}},\ldots,f^{p}_{1}}^{\alpha\text{-spin prot.}},
    \underbrace{f^{e}_{n_\beta ^{e} + n_\alpha^e },\ldots,f^{e}_{1+n_\alpha^e }}_{\beta \text{-spin elec.}},
     \overbrace{f^{e}_{n_\alpha^{e}},\ldots,f^{e}_{1}}^{\alpha\text{-spin elec.}}
    }, \;\;\;\; f_{i}^{e/p} \in \{0,1\} \, ,
\end{equation}
where $f^{e}_{i}$ and $f^{p}_{i}$ are the spin orbital occupation numbers for electrons and protons respectively, and $n_\alpha^{e}, n_\beta^{e}, n_\alpha^{p}, n_\beta^{p}$ are the number of $\alpha$ and $\beta$ spin orbitals for the two types of particle. With parity encoding, instead of storing the occupation of individual spin orbitals,
the qubits store information about the parity of the fermionic state as follows
\begin{equation}
    p^{e/p}_{i} = \bigoplus_{j=1}^{i} f^{e/p}_{i},
\end{equation}
where $\bigoplus$ represents addition modulo 2. A change to the parity encoding thus yields the parity state
\begin{equation}
    \ket{p} = \ket*{
    \underbrace{p^{p}_{n_\beta ^{p} + n_\alpha^p },\ldots,
    p^{p}_{1+n_\alpha^p}}_{\beta \text{-spin prot.}},
     \overbrace{p^{p}_{n_\alpha^{p}},\ldots,p^{p}_{1}}^{\alpha\text{-spin prot.}},
    \underbrace{p^{e}_{n_\beta ^{e} + n_\alpha^e },\ldots,p^{e}_{1+n_\alpha^e }}_{\beta \text{-spin elec.}},
     \overbrace{p^{e}_{n_\alpha^{e}},\ldots,p^{e}_{1}}^{\alpha\text{-spin elec.}}
    }, \;\;\;\; p_{i}^{e/p} \in \{0,1\} \, ,
\end{equation}
thus mapping the local state of each spin orbital to the state of several qubits. Note, that the qubit mapping for each type of particle is performed within the corresponding subspace, such that the parity of electronic qubits is not affected by the state of nuclear qubits, and vice versa. 

The fermionic creation and annihilation operators in the parity basis
have the following representations
\begin{align}
    a^{\dagger}_i &\equiv \frac{1}{2}\left[X^{\leftarrow}_{i+1} \otimes (X_i \otimes Z_{i-1} - iY_i)\right], \\
    a_i &\equiv \frac{1}{2}\left[X^{\leftarrow}_{i+1} \otimes (X_i \otimes Z_{i-1} + iY_i)\right] \, .
\end{align}
Here we introduce $X^{\leftarrow}_{i+1}$ as an operator that acts on all qubits of index $i+1$ and higher within the electronic or nuclear subspaces, accounting for fermionic antisymmetry under the exchange of identical particles. For the edge case of $i=1$, the expression simplifies to $\frac{1}{2}\left[X^{\leftarrow}_{i+1} \otimes (X_i + iY_i)\right]$, and similarly for $i = n^{e/p}_\alpha + n^{e/p}_\beta$ we get $\frac{1}{2}\left(X_i \otimes Z_{i-1} + iY_i\right)$.

The utilization of the parity encoding simplifies the identification of the discrete $\mathbb{Z}_2$-symmetries associated with the particle number operators within the $\alpha$ and $\beta$ 
 subspaces~\cite{brav2017} for both electrons and nuclei. As we are dealing with a Hamiltonian based on the non-relativistic Schr\"odinger equation, Eq.~\eqref{eq:fqSchroed}, the number of particles within the $\alpha$ and $\beta$ subspaces are conserved and therefore total parity numbers such as $p^{p}_{n_\beta ^{p} + n_\alpha^p }$, $p^{p}_{n_\alpha^p }$, $p^{e}_{n_\beta ^{e} + n_\alpha^e }$, and $p^{e}_{n_\alpha^e }$ are constant. The corresponding 4 qubits can thus be treated classically and their contribution to the expectation value of the Hamiltonian can be calculated beforehand employing the ``tapering'' procedure described in Ref.~\cite{brav2017}. 

\subsection{Qubit Reduction from Hamiltonian Symmetries}\label{sec:symred}
If a molecular system obeys some point group symmetry, the number of qubits can be further reduced using methods introduced by Bravyi \textit{et al.}~\cite{brav2017}. This approach allows for additional qubit tapering by identifying a $\mathbb{Z}_2$-symmetry of the Hamiltonian corresponding to the specific molecular point group symmetry.

In addition, when the system wave function can be restricted to specific subspaces of the total Hilbert space, one can further impose an additional $\mathbb{Z}_2$-symmetry, which confines the solutions in the region of interest.
This may, for example, happen in the case of the hydrogen molecule if one is interested in the orthohydrogen nuclear isomer, the dominant nuclear spin isomer (75\%) at room temperature. 
Thus one can just consider the maximally spin-polarized triplet state for the protons. 
This allows for either alpha or beta spin orbitals to be ignored. 
To reduce our active space to the selected set of nuclear alpha or beta orbitals, we introduce (here for the case of orthohydrogen) a projector to states for which beta spin orbitals are unoccupied
\begin{equation}\label{eq:ptriplet}
\hat P = 
\bigotimes_{J}^{\beta\text{ nuc.}} \frac{1}{2}(I+Z)
\bigotimes_I^{\alpha\text{ nuc.}}I
\bigotimes_{i}^{\text{elec.}}I
\end{equation}
and produce the projected Hamiltonian
\begin{equation}
\hat H_P = \hat P \hat H \hat P.
\end{equation}
In this case, one can find as many $\mathbb{Z}_2$-symmetries as the number of beta 
spin orbitals and the qubits corresponding to these spin orbitals can be therefore tapered off. 
This operation can also be performed in the Hamiltonian construction procedure by simply removing terms containing creation/annihilation operators acting on these spin orbitals.

\section{Results}\label{sec:res}
Our simulations focused on the evaluation of the NEO ground state energies and entanglement entropies for \acs{h2} and 
malonaldehyde. 
We first introduce a detailed description of the two systems and describe the conditions under which the conventional quantum chemistry calculations were performed. 
We then describe (in Section~\ref{sec:partran}) the procedures for the initialization of the quantum computing calculations, as well as the transformations employed in the mapping of the fermionic electronic structure problem in the qubit space (see  Sections~\ref{sec:mapping} and \ref{sec:symred}). 
In Section \ref{sec:h2}, we will report the results for the hydrogen molecule, discussing the performance of the different wavefunction ans\"atze, including the one inspired by UCC and the more quantum native, hardware efficient, TwoLocal ansatz. 
Finally, in Section~\ref{sec:malon} we will present and discuss the results for the second and more challenging system, malonaldehyde, which is a prototypical system used to investigate quantum effects in molecular proton transfer reactions. 
In this case, the delocalization of the proton wave function between the donor and acceptor moieties makes the choice of the proton basis set and wave function ansatz more challenging than in the case of \acs{h2} molecule. 

In both cases, we will also discuss the level of entanglement between the two quantum subsystems, the electronic and nuclear components, providing additional insights into the nature of the proton transfer process.

\subsection{Test Systems}\label{sec:systems}
We chose the \ac{h2} for testing the developed methodology, as the 
exact reference results are readily accessible and all possible types of 
interaction between particles such as electron-electron, proton-proton, 
and electron-proton are present. The \ac{neo} orbitals and corresponding 
integrals were prepared with the \ac{neohf} method. For ease of calculation, the 
rather small \ac{631}~\cite{631G} basis set was used for the electronic orbitals and 
the nuclear orbitals were constructed using the smallest basis set available in 
GAMESS-US~\cite{GAMESS}, \ac{dzsnb}~\cite{webb2002}. For the showcase in 
Section~\ref{sec:partran} the electronic orbitals were constructed in 
\ac{sto6}. The basis set functions for electrons and nuclei were centered at 
the positions corresponding to the experimental ground state 
structure~\cite{hube1979} with an inter-nuclear distance equal to 0.7414 \AA. We 
must stress that in general one must use the floating orbital 
approach~\cite{iord2003,hurl1988,tach1999}, where the centers of the basis set 
functions are optimized to best fit the final wave function according to the 
variational principle. However, as it entails a more elaborate 
implementation we prefer to postpone this step and perform all tests with fixed 
orbital centers, similar to the studies performed in 
Refs.~~\cite{naka2001,naka2005,bror2020,pavo2020}. Note that 
our \ac{h2} Hamiltonian, similar to the work in Ref.~\cite{tach2000}, contains contributions from global translational and rotational motion. As these contributions increase the total energy they preferably must be eliminated~\cite{kolo1963,naka2005,muol2019}. 
We emphasize that our aim is not in delivering ultimate accuracy results but rather to establish the resource-efficient methodology for the quantum treatment of the electron-nuclear problem on near-term quantum computers. 
Based on the \ac{neohf} orbitals the reference total energy has been evaluated with \ac{neofci}~\cite{webb2002} method, see 
Table~\ref{tab:uccneo}. The same structure was used for the electronic \ac{hf} 
calculation with a \ac{631} basis set giving 4 orbitals for electrons. 
Consequently, these orbitals were used in \ac{fci} calculation for obtaining 
the reference energy. The reference energies for electronic and \ac{neo} 
cases are presented in Tables~\ref{tab:uccbo} and \ref{tab:uccneo}, respectively.

\begin{figure}[htb]
    \centering
    \includegraphics{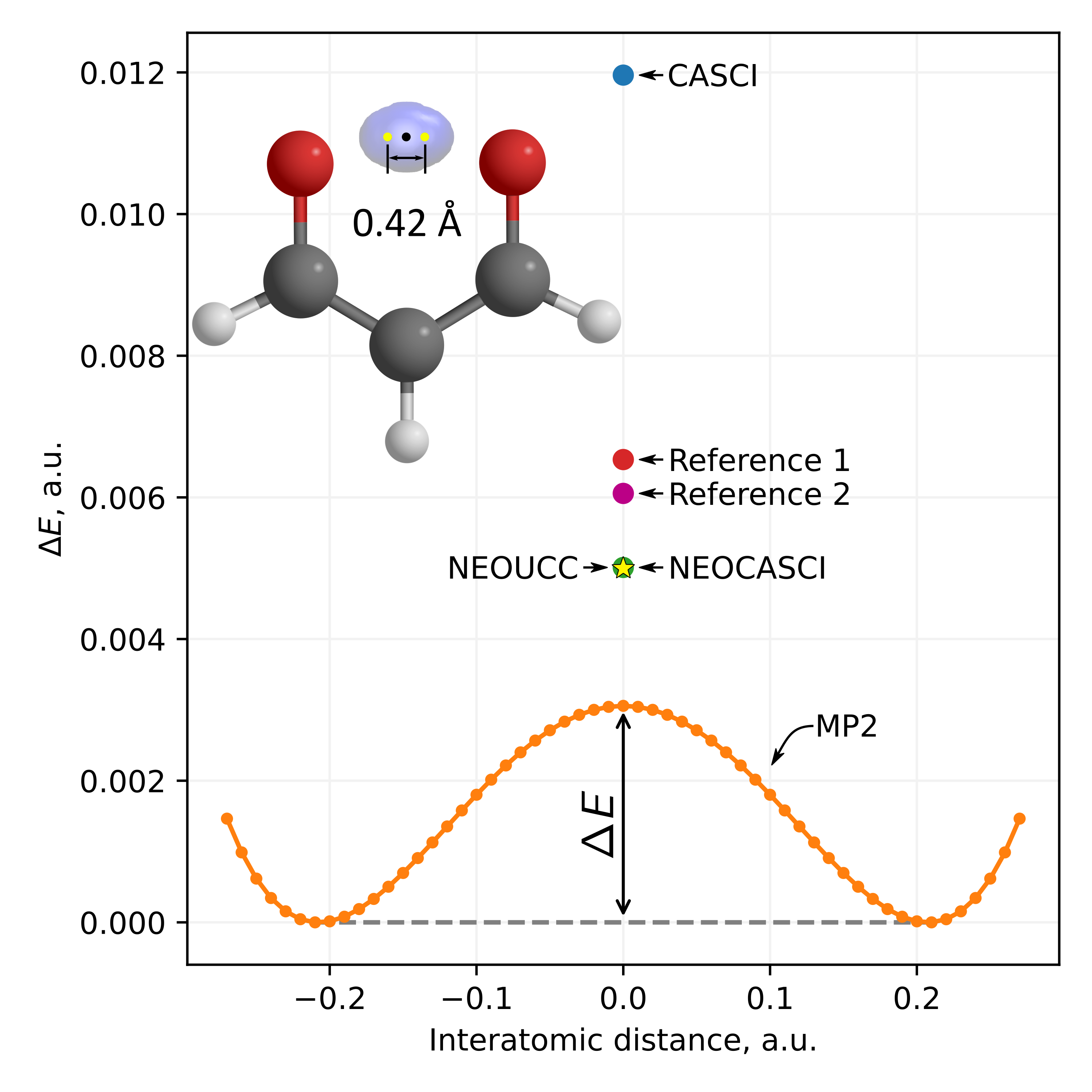}
    \caption{Potential energy surface for the proton translocation process in malonaldehyde. The curve in orange corresponds to a \acs{mp2} calculation, in which the tunneling proton (as all other atoms) is described as a classical point charge; it is given mainly as a guidance to the eyes. The path is simulated keeping all other atoms (but the moving proton) fixed in their original position. The structure of the malonaldehyde together with the first protonic \acs{neohf} orbital (isosurface value $10^{-4}$) are given in the upper inset. The distance between the equilibrium positions is found to be 0.42~\AA. Both \acs{neocasci} (classical algorithm, green dot) and \acs{neoucc} (quantum algorithm, yellow star) methods incorporate quantum proton effects and provide reasonable tunneling barrier values, while the classical description of the proton with \acs{casci} (blue dot) leads to a considerable  overestimation of the barrier. The values for ``Reference 1'' (red dot) and ``Reference 2'' (purple dot) are taken from Ref.~\cite{wang2008a} and Ref.~\cite{list2020} respectively.}
    \label{fig:malon_BO_PES}
\end{figure}

For the second test system, we chose malonaldehyde. This molecule is an extensively studied benchmark system for quantum effects involving intramolecular hydrogen bonds. The key structural feature is the hydrogen bond O–H–O for which there are two possible asymmetric configurations. This leads to a double-well potential with the states connected by proton tunneling. The experimental ground state structure~\cite{baug1981} has $C_s$ point group symmetry and the proton transfer occurs at a symmetric transition state structure in $C_{2v}$ symmetry~\cite{flud1977,webb2002}. To get a reasonable system for the description of the tunneling phenomena we first optimized the malonaldehyde structure using \ac{mp2} method~\cite{moll1934} in $C_{2v}$ using the \ac{631} basis set. The optimized structure corresponds to the \ac{ts} structure~\cite{flud1977,webb2002}. After fixing all the coordinates but one, the search for minima was performed in $C_s$ symmetry. This revealed two possible equilibrium locations for the proton separated from each other by 0.42 {\AA} as well as the location of the barrier's maximum. In Figure~\ref{fig:malon_BO_PES} one can see the equilibrium positions marked with yellow dots symmetrically located with respect to the position of the barrier maximum marked with a black dot. 
These coordinates were used as the centers for the nuclear basis functions for three different \ac{neohf} calculations with \ac{dzsnb} and \ac{631} basis sets for nuclei and electrons respectively. Two \ac{neohf} calculations were performed in $C_{2v}$ symmetry representing \ac{ts} place of nuclear orbitals (i) at equilibrium positions and (ii) at equilibrium together with the barrier's maximum positions. The former will be referred to as ``2-center $C_{2v}$'' and has 4 nuclear orbitals, while the latter, ``3-center $C_{2v}$'', has 6 nuclear orbitals.

\begin{figure}[htb]
    \centering
    \includegraphics[width=0.32\textwidth]{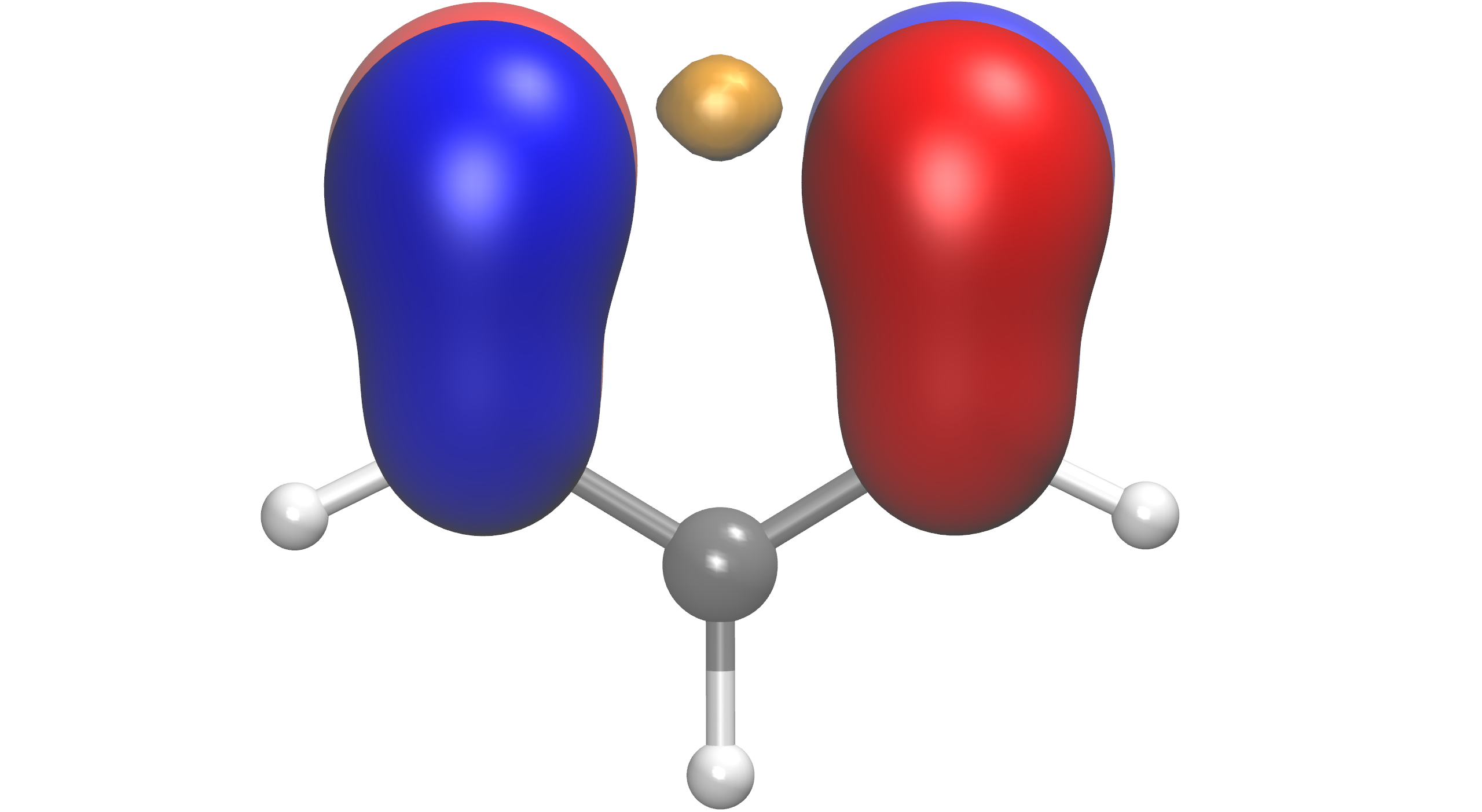}
    \includegraphics[width=0.32\textwidth]{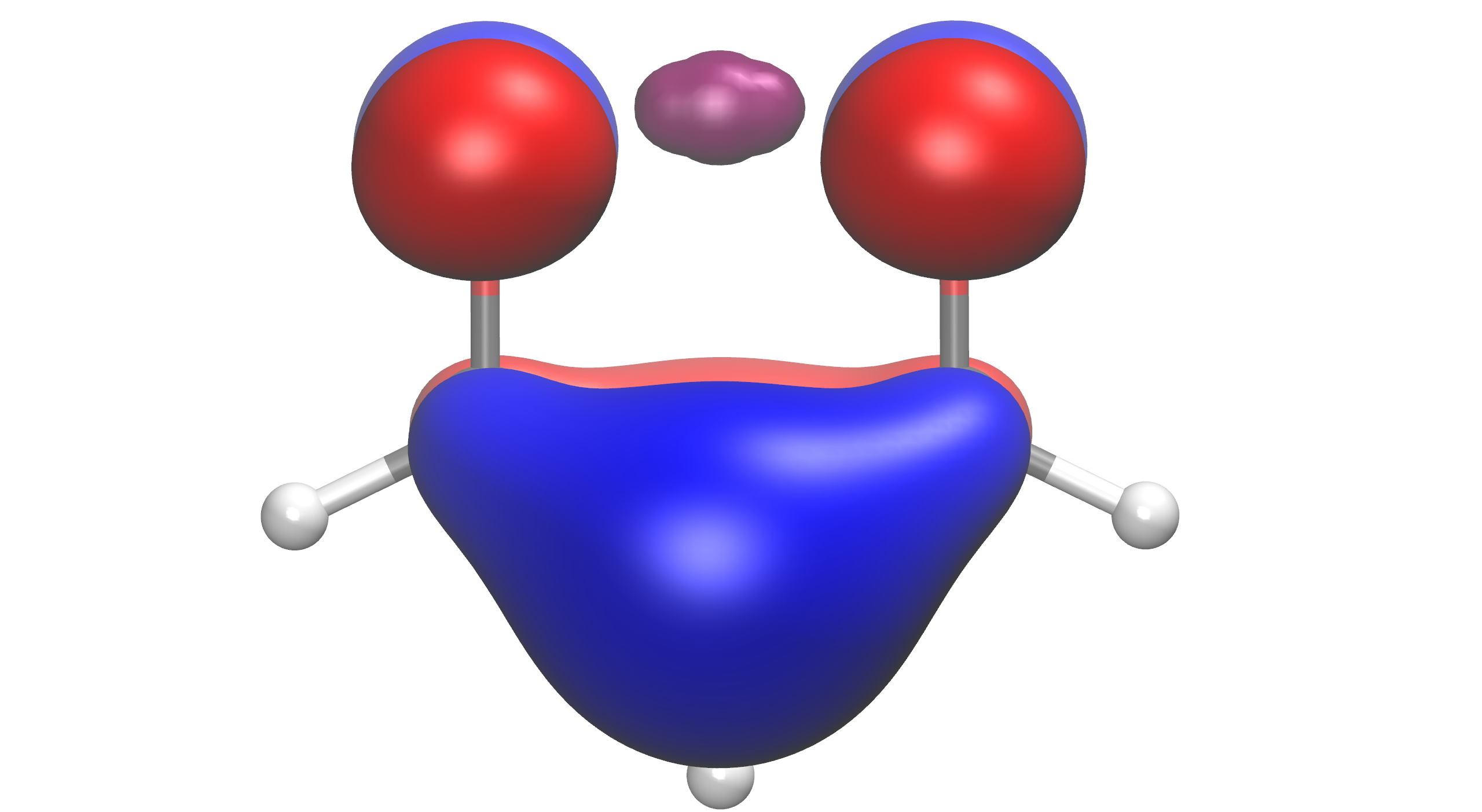}
    \includegraphics[width=0.32\textwidth]{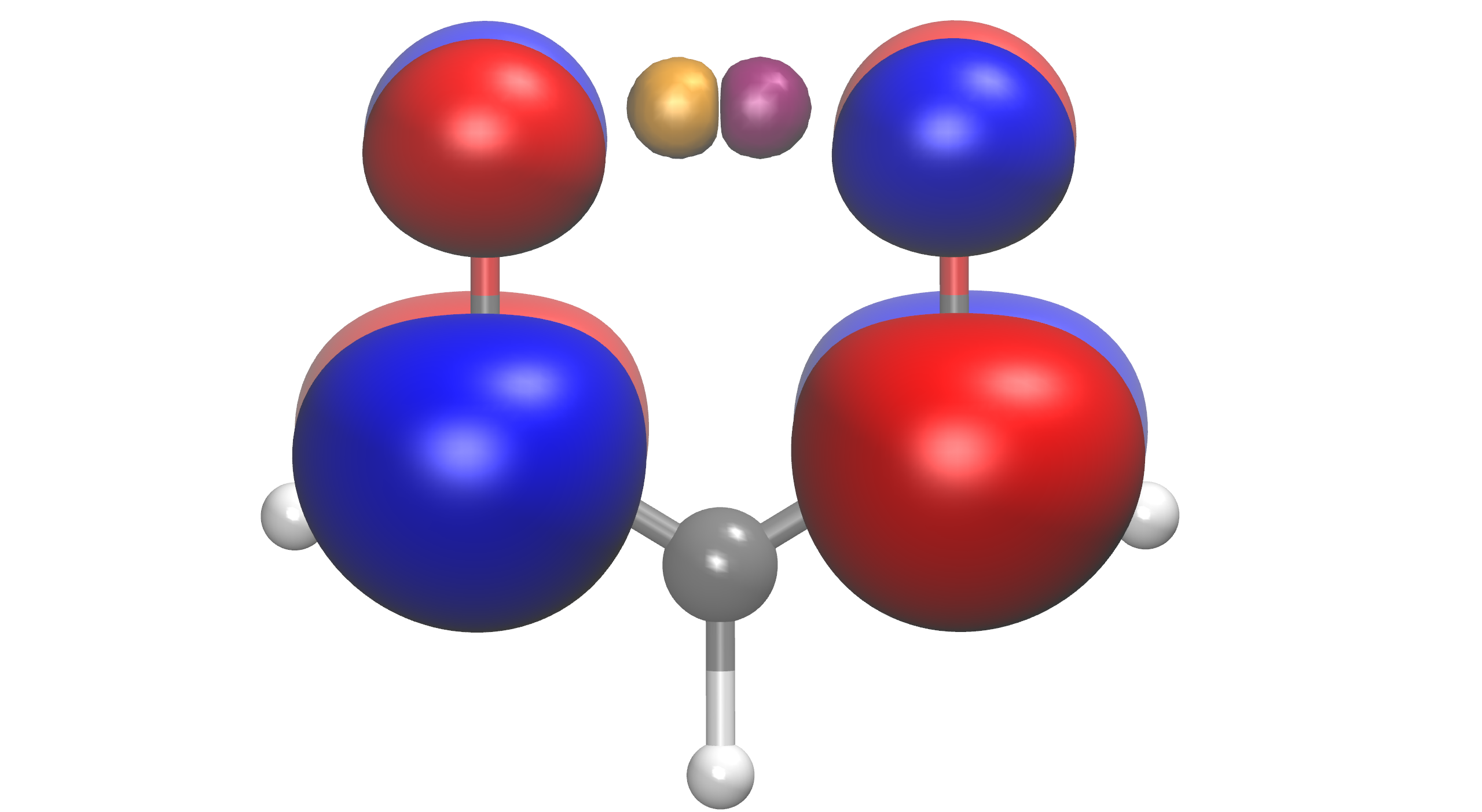}
    \includegraphics[width=0.32\textwidth]{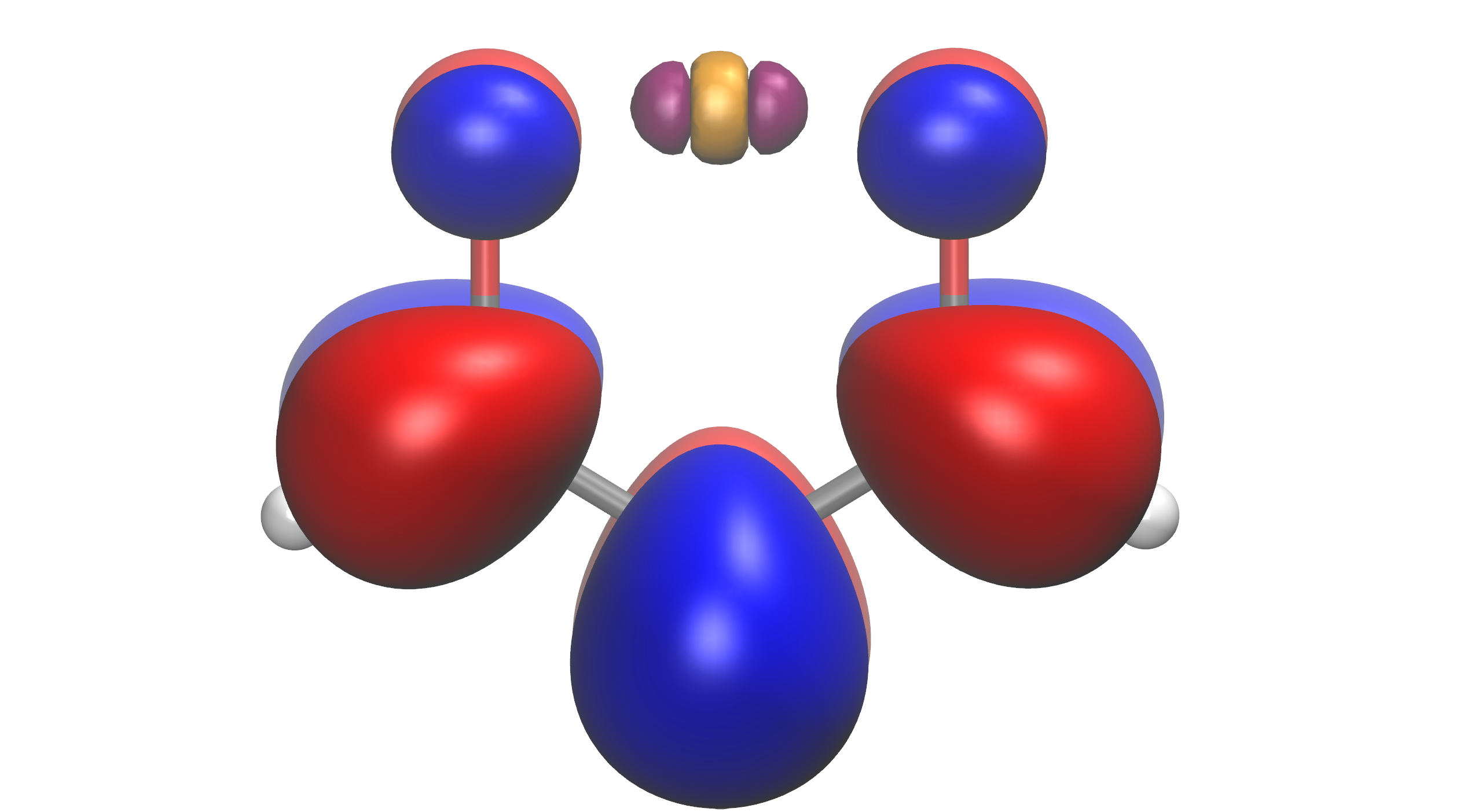}
    \includegraphics[width=0.32\textwidth]{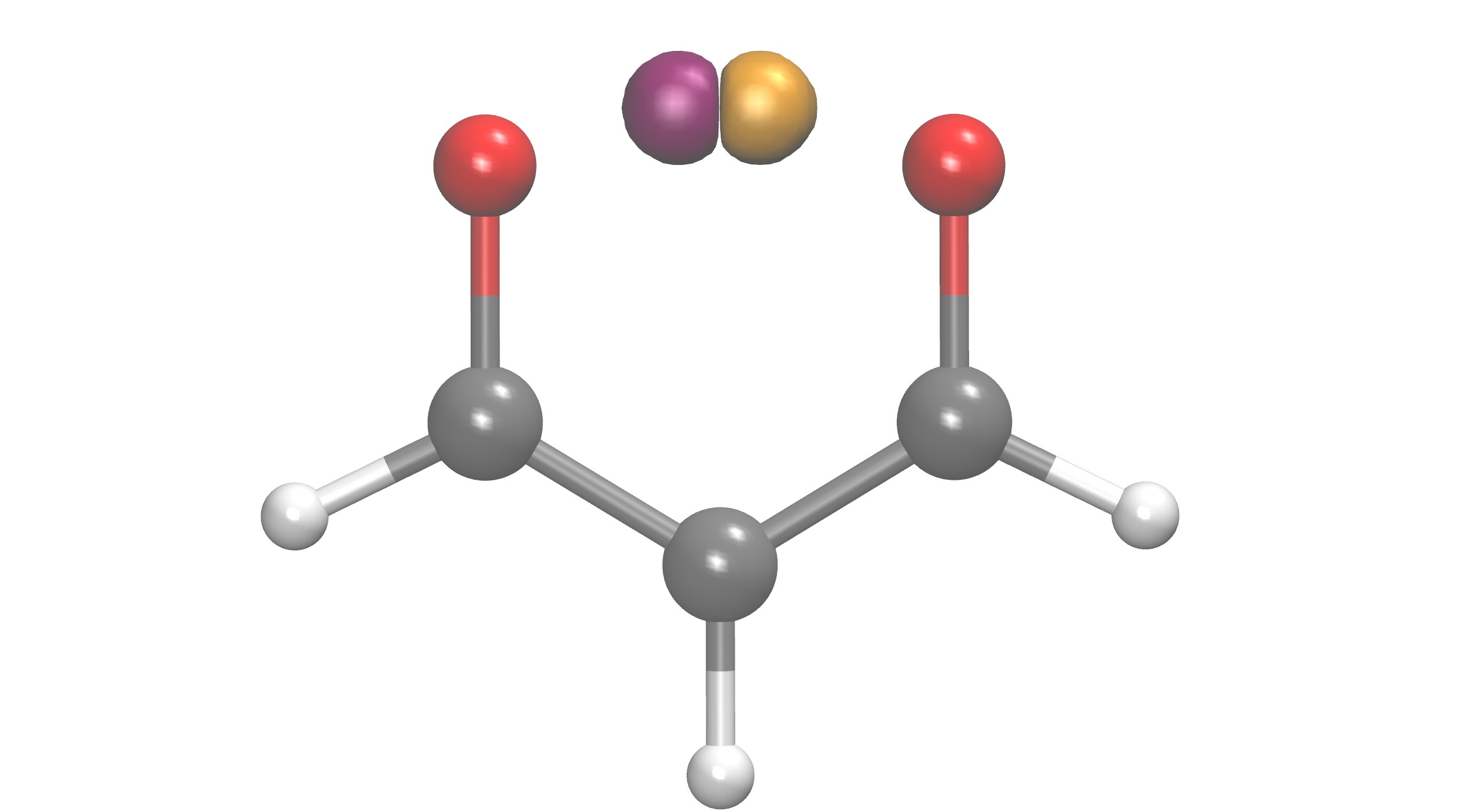}
    \includegraphics[width=0.32\textwidth]{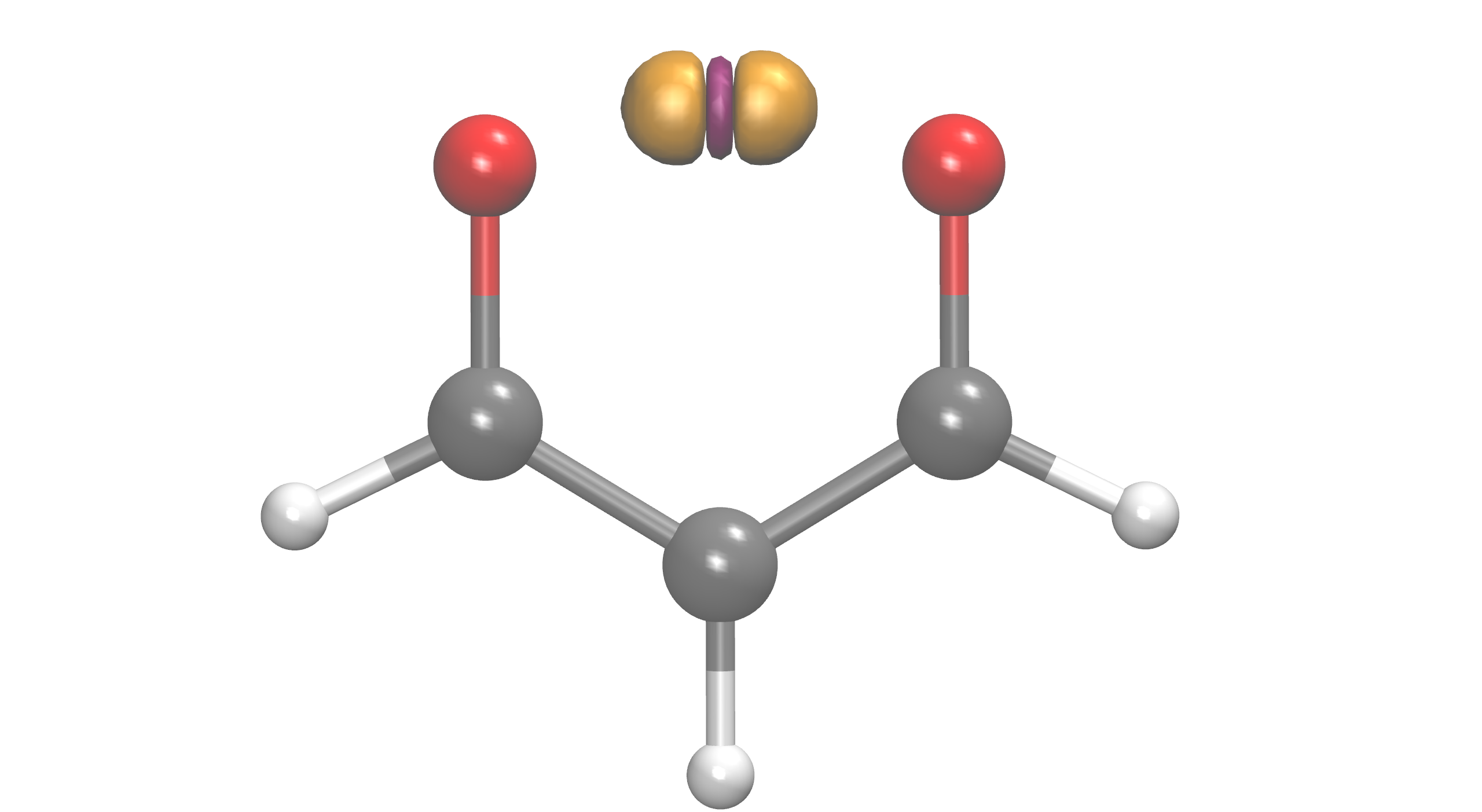}
    \caption{Electronic (red/blue) and nuclear (purple/orange) orbitals (isosurface value is $0.05$) included in the active space of \ac{neocasci} calculation for malonaldehyde in the ``3-center $C_{2v}$'' setup. The orbitals have been prepared with a \ac{neohf} calculation using a \ac{dzsnb} and a \ac{631} basis sets for the nuclei and the electrons, respectively.}
    \label{fig:malon_orbitals}
\end{figure}

The third \ac{neohf} calculation represents the ground state, ``1-center $C_s$'', places nuclear orbitals at one of the equilibrium locations and has only 2 nuclear orbitals. In all cases, only one proton was treated quantum mechanically, while all other nuclei were taken as classical point charges. We also kept the same number of electronic orbitals resulting from the tunneling proton, placing them on all three centers (ghost atoms feature) for achieving an equal description of the electronic wave function in case of ``1-center $C_s$'' and ``3-center $C_{2v}$'' setups. Due to the presence of the classical nuclei, the \ac{neo} Hamiltonian does not contain the global rotational and translational degrees of freedom~\cite{yang2019}. 
Based on the three sets of \ac{neohf} orbitals (``1-center $C_s$'', ``2-center $C_{2v}$'', ``3-center $C_{2v}$'') we have performed \ac{neocasci} calculations in order to obtain the corresponding reference energies, see Table \ref{tab:Malon}. To keep the electronic active space compact the 17 low-lying core orbitals for electrons were considered fully occupied and only 4 orbitals hosting 4 electrons were considered active in all calculations. For the nuclear active space, we used all 2 orbitals, 4 orbitals, and 6 orbitals for  ``1-center $C_s$'', ``2-center $C_{2v}$'', and ``3-center $C_{2v}$'' setups respectively. The reference results for \ac{neocasci} are presented in Table~\ref{tab:Malon} while the electronic and nuclear \ac{neohf} orbitals used in the active space for the \ac{neocasci} calculation on ``3-center $C_{2v}$'' setting are presented in Figure~\ref{fig:malon_orbitals}. 

\subsection{Initialization of the Electronic Parameters in \ac{neo} Ansatz}\label{sec:partran}

As discussed in Section~\ref{sec:ansatzinit}, the choice of initial parameters in any VQE ansatz can have a significant influence on the convergence rate as well as the accuracy of a calculation. We show here that the optimized wave function of the electronic subsystem offers a good starting point to the \ac{neo} algorithm. As was already mentioned earlier the transfer of parameters between the electronic and \ac{neo} ans{\"a}tze might need additional modifications if a point group symmetry is manifest. 
To this end, one needs to consider the different sign patterns occurring in the \ac{neo} and electronic Hamiltonians after applying the $\mathbb{Z}_2$-symmetry reduction corresponding to a certain point group symmetry. 
We demonstrate this procedure on the example of the hydrogen molecule using \ac{neohf} orbitals constructed with minimal size \ac{sto6} and \ac{dzsnb} basis sets for electrons and nuclei respectively as explained in Section~\ref{sec:systems}. Thus we keep the problem size small for illustration purposes and deal only with 4 orbitals for nuclei and 2 orbitals for electrons. This results in a \ac{neo} qubit Hamiltonian spanning 12 spin orbitals (8 nuclear spin orbitals and 4 electronic spin orbitals), 
while the electronic Hamiltonian spans just 4 spin orbitals. 
The $\mathbb{Z}_2$-symmetry arising from the $D_{2h}$ point group symmetry of the electronic qubit Hamiltonian in parity mapping is given by the following operator
\begin{equation}
\mathbb{Z}_2^{\text{El.}} = IZIZ\,.
\end{equation}
Applying tapering to the electronic Hamiltonian leads to a projection into the symmetry subspace corresponding to the HF state with $\mathbb{Z}_2^{\text{El.}}$ eigenvalue equal to -1. Thus, after two-qubit reduction~\cite{brav2017} and symmetry tapering the resulting electronic Hamiltonian will take the following form
\begin{align}\label{eq:1qbo}
\hat H_{\text{El.}} =  
-0.322833\times I  -0.803007\times Z -0.180939\times X\, .
\end{align}
For the \ac{neo} qubit Hamiltonian in parity encoding the corresponding $\mathbb{Z}_2$-symmetry is
\begin{equation}
\mathbb{Z}_2^{\text{\ac{neo}}} = IZZZIZZZIZIZ \, .
\end{equation}
However, in this case, the symmetry operator eigenvalue is +1. 
Thus the eigenvalue with a different sign will be incorporated into the Hamiltonian during the symmetry tapering. To reveal this effect, we further perform 4-qubit reduction according to Section~\ref{sec:mapping} and project the \ac{neo} Hamiltonian onto the electronic subspace. The corresponding projector, $P_e$, is similar to the one introduced in Section~\ref{sec:symred} except that it additionally projects the \ac{neo} qubit Hamiltonian onto states where the nuclear alpha 
spin orbitals occupied according to the lowest \ac{neohf} occupation. 
Thus the resulting \ac{neo} qubit Hamiltonian,
\begin{equation}\label{eq:1qneo}
\hat P_e \hat H_{\text{\ac{neo}}} \hat P_e =
-0.217313\times I -0.816590\times Z +0.181576\times X\,,
\end{equation}
is limited to the states where only the energetically lowest triplet \ac{neohf} occupation for protons is allowed. One can clearly recognize in Eq.~\eqref{eq:1qneo} that the off-diagonal term acquires a change of sign when compared to the electronic case, Eq.~\eqref{eq:1qbo}. This difference in the sign signature is also present for the parameters in the electronic and \ac{neo} ans\"atze. 
One way to solve this issue would be to perform a signature similarity transformation of the electronic Hamiltonian. For  the specific case discussed here, this will correspond to
\begin{equation}\label{eq:1qzbo}
Z \hat H_{\text{El.}} Z = 
-0.322833\times I  -0.803007\times Z +0.180939\times X \, .
\end{equation}
Since such a transformation is isospectral~\cite{char2013}, one can use the transformed electronic Hamiltonian to find the parameters of the electronic ansatz, and then apply these parameters as an initial guess for the \ac{neo} ansatz. 
In this work, this ``signature similarity transformation'' is performed by comparing and fixing the signs of the Pauli words of the projected \ac{neo} Hamiltonian and of the electronic Hamiltonian according to the expressions in Eq.~\eqref{eq:1qbo} and Eq.~\eqref{eq:1qneo}.

\begin{figure}[htb]
{\small
\begin{equation*}
    {\centering
    \Qcircuit @C=0.5em @R=.7em {
    &&&&\text{electronic part}&&&&&\\
    & \lstick{q_{0}} & \gate{\tB{R_y(\theta_1)}} & \gate{\tB{R_z(\theta_2)}} & \ctrl{1}     & \ctrl{2}     & \ctrl{3}     & \gate{\tR{R_y(\tht{d,1})}} & \gate{\tR{R_z(\tht{d,2})}}
    & \qw\gategroup{2}{3}{2}{9}{.5em}{--} \\
    & \lstick{q_{1}} & \gate{\tR{R_y(\tht{1})}}      & \gate{\tR{R_z(\tht{2})}}      & \control \qw & \qw          & \qw          & \gate{\tR{R_y(\tht{d,3})}} & \gate{\tR{R_z(\tht{d,4})}} & \qw \\
    & \lstick{q_{2}} & \gate{\tR{R_y(\tht{3})}}      & \gate{\tR{R_z(\tht{4})}}      & \qw          & \control \qw & \qw          & \gate{\tR{R_y(\tht{d,5})}} & \gate{\tR{R_z(\tht{d,6})}} & \qw \\
    & \lstick{q_{3}} & \gate{\tR{R_y(\tht{5})}}      & \gate{\tR{R_z(\tht{6})}}      & \qw          & \qw          & \control \qw & \gate{\tR{R_y(\tht{d,7})}} & \gate{\tR{R_z(\tht{d,8})}} &
    \qw \gategroup{5}{5}{5}{9}{0.8em}{_\}}\\
    \\
    & & & & & & &\text{~~~~$d$ layers} & &
    }
}
\end{equation*}
}
\caption{The ``stacked'' variant for the initialization of the electronic parameters in hardware-efficient ansatz. The circuit is designed for the case of 4-qubit \ac{neo} Hamiltonian, corresponding to \ac{h2} using minimal basis sets, \ac{sto6} for electrons and \ac{dzsnb} for nuclei. The electronic and nuclear parameters are shown in blue and red respectively.
In this case, entangling CZ gates are only applied between electronic and nuclear subsystem qubits. 
The electronic subsystem is delimited by a dashed box. 
Parameters in blue are initialized according to the results of the electronic subsystem optimization.
All parameters in red: $\tht{i}$ ($i=1,6$) and $\tht{d,i}$ ($i=1,8$) are initialized to 0, with the exception of  $\tht{1}$, 
which is set to $\pi$. 
Note that the number of layers, $d$, should be even in order for the electronic parameters to remain as close as possible to their initial values (\textit{i.e.}, for avoiding the phase flip induced by the CZ gates). }
\label{circ:TLst}
\end{figure}
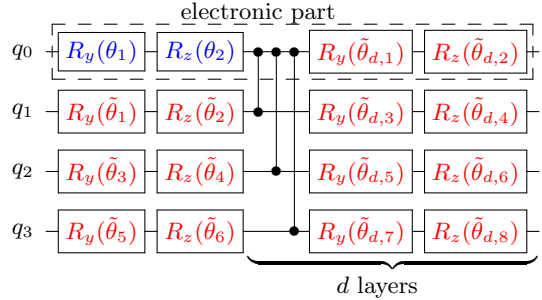

The similar weights (coefficients) obtained for the projected \ac{neo} and in the electronic Hamiltonians, Eq.~\eqref{eq:1qneo} and Eq.~\eqref{eq:1qzbo}, attests to the efficacy of the proposed parameter initialization methodology. 
However, before transferring electronic parameters to the \ac{neo} ansatz, it could be beneficial to additionally optimize them using the projected \ac{neo} Hamiltonian $\hat P_e \hat H_{\text{\ac{neo}}} \hat P_e$ (Eq.~\eqref{eq:1qneo}). 
More details will be given in Section \ref{sec:h2}. 
In the following, 
we limit ourselves to the case of the initialization of the electronic parameters for the hardware-efficient ansatz for \ac{h2} in the minimal basis set. 
After performing symmetry tapering, 4-qubit reduction, and restricting to nuclear triplet states according to Sections~\ref{sec:mapping} and \ref{sec:symred}, one finally obtains a 4-qubit \ac{neo} Hamiltonian (See Appendix \ref{sec:minimalH2}, Eq.~\eqref{eq:minimalH2}). 
The first qubit (of the 4-qubit register) corresponds to the electronic state, while the other 3 qubits describe the nuclear state (see Figure~\ref{circ:TLex}). 
One can initialize electronic parameters simply by transferring the wave function of the electronic subsystem into the \ac{neo} ansatz, as shown in  Figure~\ref{circ:TLex}. 
All nuclear parameters must then be initialized to zero except for the second $R_y$ gate on qubit $q_{1}$. 
In this specific case, it is set to $\pi$ resulting in the lowest energy triplet nuclear occupation. This procedure, that we will denote ``expanded'', can be easily generalized to larger system sizes. 
A second more economic variant only requires the additions of entangling gates between the electronic and the nuclear qubit subregisters, as shown in Figure~\ref{circ:TLst}. 
This variant, referred to as ``stacked'', can be also generalized to larger systems and is advantageous in cases with weak entanglement between the protons. 
It is worth mentioning that for the electronic parameters (in blue color in Figure~\ref{circ:TLst}) to remain as close as possible to their initial values (\textit{i.e.}, for avoiding the phase flip induced by the CZ gate) the number of entangling layers is required to be even.
\begin{figure}[htb]
{\small
\begin{equation*}
{\centering
\Qcircuit @C=0.5em @R=.7em {
&&&& \text{electronic part} &&&&&&&&\\
&\lstick{q_{0}}&\gate{\tB{R_y(\theta_1)}}&\gate{\tB{R_z(\theta_2)}}&\ctrl{1}    &\ctrl{2}    &\qw         &\ctrl{3}    &\qw         &\qw         &
\gate{\tB{R_y(\theta_3)}} & \gate{\tB{R_z(\theta_4)}} &\qw 
\gategroup{2}{3}{2}{12}{.5em}{--}\\
&\lstick{q_{1}}&\gate{\tR{R_y(\tht{1})}}     &\gate{\tR{R_z(\tht{2})}}     &\control \qw&\qw         &\ctrl{1}    &\qw         &\ctrl{2}    &\qw         &
\gate{\tR{R_y(\tht{d,1})}} & \gate{\tR{R_z(\tht{d,2})}} &\qw \\
&\lstick{q_{2}}&\gate{\tR{R_y(\tht{3})}}     &\gate{\tR{R_z(\tht{4})}}     &\qw         &\control \qw&\control \qw&\qw         &\qw         &\ctrl{1}    &
\gate{\tR{R_y(\tht{d,3})}} & \gate{\tR{R_z(\tht{d,4})}} &\qw \\
&\lstick{q_{3}}    &\gate{\tR{R_y(\tht{5})}} &\gate{\tR{R_z(\tht{6})}}     &\qw         &\qw         &         \qw&\control \qw&\control \qw&\control \qw&
\gate{\tR{R_y(\tht{d,5})}}&\gate{\tR{R_z(\tht{d,6})}} &\qw \gategroup{5}{5}{5}{12}{0.8em}{_\}}\\
\\
&&&&&&&&&\mbox{~~~~~~~~~~~~~~~$d$ layers}\\
}
}
\end{equation*}
}
\caption{The ``expanded'' variant for the initialization of the electronic parameters in hardware-efficient ansatz. The circuit is designed for the case of 4-qubit \ac{neo} Hamiltonian, corresponding to \ac{h2} using minimal basis sets, \ac{sto6} for electrons and \ac{dzsnb} for nuclei. We use an ``all-to-all'' entangling scheme with CZ gates applied between all qubits. 
The electronic subsystem is delimited by a dashed box. 
Parameters in blue are initialized according to the results of the electronic subsystem optimization.
All parameters in red: $\tht{i}$ ($i=1,6$) and $\tht{d,i}$ ($i=1,6$) are initialize to 0, with the exception of  $\tht{d_{\text{max}},1}$,  which is set to $\pi$ ($d_{max}$ is the maximum number of repetitions).
While for this specific case one entangling layer, $d=1$, would be sufficient, for larger systems more layers will be required. }
\label{circ:TLex}
\end{figure}
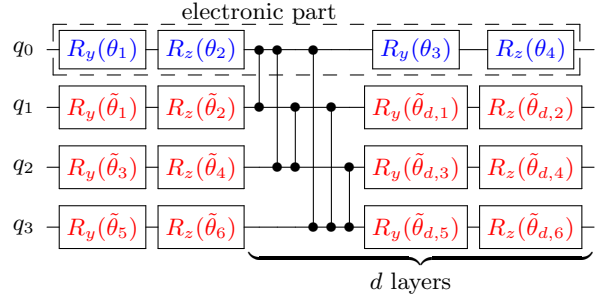

\subsection{\acs{h2} calculations}\label{sec:h2}

Initially, we have performed \ac{vqe} optimization of the electronic \ac{uccsd} ansatz for the ground state of the \ac{h2} molecule employing an electronic Hamiltonian with classical nuclei, $\hat{H}_{\text{El.}}$. The final energy agrees well with the reference value obtained from  \ac{fci} calculations, see Table~\ref{tab:uccbo}. To verify the validity of the electronic \ac{uccsd} parameters for the \ac{neo}\ac{uccsd} ansatz we employ the projected \ac{neo} Hamiltonian, $P_e\hat{H}_{\ac{neo}}P_e$, in a manner similar to Eq.~\eqref{eq:1qneo} in Section~\ref{sec:partran}. After projection and tapering, the 5-qubit \ac{neo} 
Hamiltonian considers only states where the nuclear part corresponds to the lowest 
energy triplet nuclear \ac{neohf} occupation and has the same size as the electronic qubit Hamiltonian. The corresponding ground state energy of 1.065040 Hartree has been calculated by exact diagonalization of $P_e\hat{H}_{\text{\ac{neo}}}P_e$. Initializing the electronic parameters in \ac{neo}\ac{ucc}S$^{(1,0)}$D$^{(2,0)}$ (see notation in Section~\ref{sec:ansaetze}) to the values from electronic \ac{uccsd} ansatz, we have evaluated the corresponding expectation value, $\langle P_e\hat{H}_{\text{\ac{neo}}}P_e \rangle$. One can see from Table~\ref{tab:uccbo} that the resulting expectation value approximates the ground state reference energy for $P_e\hat{H}_{\text{\ac{neo}}}P_e$ within an error of less than $10^{-4}$ Hartree. By letting the electronic parameters further relax in optimization employing the \ac{neo}\ac{ucc}S$^{(1,0)}$D$^{(2,0)}$ ansatz and  $P_e\hat{H}_{\text{\ac{neo}}}P_e$ one removes this error, recovering the exact reference energy, see Table~\ref{tab:uccbo}. 
\begin{table}[htb]
\caption{\ac{h2} ground state energies calculated with quantum computing ans\"atze (\ac{uccsd} and \ac{neo}\ac{ucc}S$^{(1,0)}$D$^{(2,0)}$) and classical algorithms (\ac{fci} and ``exact diagonalization''). Two Hamiltonians are used: the electronic Hamiltonian ($\hat{H}_{\text{El.}}$) and the projected \ac{neo} Hamiltonian ($P_e\hat{H}_{\text{\ac{neo}}}P_e$). 
The \ac{uccsd} electronic parameters reproducing the \ac{fci} wave function provide a good initial guess for \ac{neo}\ac{ucc}S$^{(1,0)}$D$^{(2,0)}$ ansatz (applied on $P_e\hat{H}_{\text{\ac{neo}}}P_e$) with an error of $8.35 \times 10^{-5}$ Ha Further optimization of parameters (``relaxed'') recovers the ground state energy of the $P_e\hat{H}_{\text{\ac{neo}}}P_e$ (``exact diagonalization'').}
\label{tab:uccbo}
\begin{center}
\begin{tabular}{llcr}
\toprule
Hamiltonian & Method    & Parameters & Energy/Ha \\
\midrule
$\hat{H}_{\text{El.}}$ & \acs{fci}     &                     & -1.151683 \\ 
                                & \ac{uccsd}    & electronic & -1.151683 \\
\midrule
$P_e\hat{H}_{\text{\ac{neo}}}P_e$ & exact diagonalization &          & -1.065040 \\
\cmidrule{2-4}
& \ac{neo}\ac{ucc}S$^{(1,0)}$D$^{(2,0)}$ & electronic         & -1.064956 \\
                                       & & relaxed & -1.065040 \\ 
\bottomrule
\end{tabular}
\end{center}
\end{table}
Having resolved the reduced problem based on $P_e\hat{H}_{\text{\ac{neo}}}P_e$ Hamiltonian we proceed to the full 8-qubit \ac{neo} Hamiltonian. After symmetry tapering, 4-qubit reduction, and projection to the nuclear triplet state according to Sections~\ref{sec:symred} and \ref{sec:mapping}, the electronic state is spanned by 5 qubits, and 3 qubits are used for the nuclear state. The results for the \ac{vqe} energy optimizations employing various types of \ac{neo}\ac{ucc} parameterizations with advanced and ordinary parameter initialization are presented in Table \ref{tab:uccneo}. For the advanced initialization procedure, we employ relaxed electronic parameters obtained with \ac{neo}\ac{ucc}S$^{(1,0)}$D$^{(2,0)}$ and $P_e\hat{H}_{\text{\ac{neo}}}P_e$ as discussed above. All variants of the \ac{neo}\ac{ucc} parameterization in Table \ref{tab:uccneo} contain double and single excitations for nuclei and electrons, which will be indicated as
\begin{equation}
\text{SD} \equiv \text{S}^{(1,0)}\text{S}^{(0,1)}\text{D}^{(0,2)}\text{D}^{(1,1)}\text{D}^{(2,0)}\, .
\end{equation}
If mixed electronic-nuclear double excitations are added then the resulting \ac{neo}\ac{ucc}DD$^{(1,1)}$ ansatz deliver the results with an error below $10^{-4}$ Hartree. The addition of the mixed triple, T$^{(1,2)}$, excitations in the \ac{neo}\ac{ucc}SDD$^{(1,1)}$T$^{(1,2)}$ ansatz does not decrease the error further. If instead, one adds the mixed triple excitation with double electronic excitation, T$^{(2,1)}$, the error decreases below $5\times10^{-6}$ Hartree. Including both variants of mixed triple excitation brings the error below $10^{-6}$
Hartree. It is clear that the D$^{(1,1)}$ and the T$^{(2,1)}$ operators are the primary contributors to the lowering of the energy, while the T$^{(1,2)}$ operator has minimal effect. 
\begin{table}[htb]
\caption{Ground state energies for the \ac{h2} \acs{neo} Hamiltonian 
obtained with different variants of the \ac{neo}\ac{ucc} quantum computing ansatz 
and the \ac{cobyla} optimizer. In general, a good agreement with \ac{neofci} is reached for all variants, however, one should emphasize the importance of the T$^{(2,1)}$ excitations in delivering excellent agreement with the reference. The calculations employ both ``ordinary'' and ``advanced'' initialization procedures, as described in Sec.~\ref{sec:partran}. Although the ``advanced'' initialization scheme improves slightly the accuracy, we expect more significant improvements in the case of a larger number of parameters.}
\label{tab:uccneo}
\begin{center}
\begin{tabular}{@{}lll@{\hskip 5mm}rr@{}}
\toprule
& \multicolumn{2}{c}{Method} & Energy/Ha & iterations\\
\cmidrule(r){1-3}
& Order &      Initialization     &  &\\
\midrule
\multirow{8}{*}{\rotatebox[origin=c]{90}{\ac{neo}\ac{ucc}}}
&               SD      & advanced &-1.066039  & 856\\
                      & & ordinary &-1.066039  & 1341\\
\cmidrule(l){2-5}
&           SDT$^{(1,2)}$ & advanced & -1.066039  & 1093\\
                      & & ordinary & -1.066039  & 1674\\
\cmidrule(l){2-5}
           & SDT$^{(2,1)}$ & advanced & -1.066120 & 1377 \\
           &             & ordinary & -1.066121 & 2179 \\
\cmidrule(l){2-5}
& SDT & advanced &-1.066121  & 1074 \\
             &           & ordinary & -1.066120 & 3945 \\
\midrule
 & \multicolumn{1}{l}{\ac{neofci}} & & -1.066121 & \\
\bottomrule
\end{tabular}
\end{center}
\end{table}
The difference in final \ac{neo}\ac{ucc} energy between optimizations with 
advanced initialization (set to relaxed electronic parameters) and ordinary 
initialization (set to the \ac{neohf} state) fall close to the convergence tolerance, $\epsilon \leq 10^{-6}$ Hartree. For \acs{neoucc} ans\"atze the advanced initialization scheme improves accuracy for the cases with a large number of variational parameters such as \ac{neo}\ac{ucc}SDT. However, the main benefit lies in a faster convergence rate, in certain cases reaching a 4-fold speedup, see Table~\ref{tab:uccneo}. Additional analysis for convergence of \ac{neo}\ac{ucc} ansatz using different initialization schemes can be found in Appendix~\ref{sec:PTconvergence}.

Table~\ref{tab:ucch2dec} presents the assessment for the contribution 
of each type of excitation operator to the correlation energy in 
\ac{neo} wave function. The correlation energy within the chosen 
basis set amounts to 0.024634 Hartree. If one uses double and single excitation operators for the protons only then energy can not decrease lower 
than \ac{neohf}. Adding the single and double operators, \ac{neo}\ac{ucc}SD, 
for electrons brings the energy nearly to the same value as for the 
case when only electronic operators are included, see Table 
\ref{tab:ucch2dec}. This can be attributed to the low correlation 
between protons. However, the mixed electron-nuclear operators make 
a substantial contribution to the correlation energy. Specifically, 
addition of mixed double excitation, D$^{(1,1)}$, brings error below
$10^{-4}$ Hartree. If one also includes mixed triple operators, T$^{(2,1)}$,
the error falls further below $5\times10^{-6}$ Hartree. The ultimate 
accuracy of $5\times10^{-7}$ Hartree is reached if all excitation 
operators up to triple-order are included. 
\begin{table}[htb]
    \caption{The ground state energies for the \ac{h2} \acs{neo} Hamiltonian obtained with different variants of \ac{neo}\ac{ucc} quantum computing ansatz and classical computational methods (\ac{neohf} and \ac{neofci}). Increasing the excitation order in the \ac{neo}\ac{ucc} ansatz lowers the error in the calculation. The mixed D$^{(1,1)}$ nuclear-electron operators have a larger contribution to the correlation than T operators, while the mixed triple operators with higher electronic excitation order, T$^{(2,1)}$, play a more important role than T$^{(1,2)}$. The significance of pure nuclear operators is minimal. 
    }\label{tab:ucch2dec}
    \begin{center}
        \begin{tabular}{@{}llrr@{}}
        \toprule
         & Method                                         & Energy/Ha   & Error/Ha\\
        \midrule
         &\ac{neofci}                                    & -1.066121 &             \\
         &\ac{neohf}                                     & -1.041487 & 0.024634\\
        \midrule
         &S$^{(0,1)}$D$^{(0,2)}$                         & -1.041487 & 0.024634\\
         &S$^{(1,0)}$D$^{(2,0)}$                         & -1.065040 & 0.001082\\
         &S$^{(1,0)}$D$^{(2,0)}$D$^{(1,1)}$              & -1.066037 & 0.000084\\
         &S$^{(1,0)}$D$^{(2,0)}$T$^{(2,1)}$              & -1.065063 & 0.001059\\
         &S$^{(1,0)}$D$^{(2,0)}$D$^{(1,1)}$T$^{(2,1)}$   & -1.066117 & 0.000004\\
         &SD$^{(2,0)}$D$^{(0,2)}$                        & -1.065049 & 0.001073\\
         &SDT$^{(2,1)}$                                  & -1.066121 & 0.000001\\
 \multirow{-9}{*}{\rotatebox[origin=c]{90}{\ac{neo}\ac{ucc}}}
         &SDT                                            & -1.066121 & 0.000001\\
        \bottomrule
        \end{tabular}
    \end{center}
\end{table}

We supplement our study of the correlation nature in the \ac{h2} molecule with the measure of entanglement entropy. Specifically, we evaluate the entanglement between electronic and nuclear subsystems, according to Section~\ref{sec:qi}. We find that the entanglement for the \ac{h2} molecule at equilibrium based on the \ac{neofci} wave function amounts to 0.0069. The same value can be accurately reproduced with the \acs{neoucc} ansatz (see Table~\ref{tab:h2_von_neumann}) when quadruple excitation operators are included. To monitor the change in entanglement upon dissociation, we also evaluated \ac{neofci} entropies for distances of 1.9582~{\AA} and 3.1751~{\AA}. 
Note that these quantities correspond to the separation between the centers hosting the nuclear and electronic orbitals and are taken as an approximated measure of the inter-atomic distance. 
Table~\ref{tab:h2_von_neumann} shows how the entanglement between electron and proton subsystems (according to \ac{neofci}) strengthens to 0.0103 during the ``bond extension'' and then slightly declines to 0.0097 after ``dissociation''. The corresponding \acs{neoucc} results are in good agreement with the reference calculations, demonstrating the reliability of the 
\acs{neoucc} approach.

\begin{table}[htb]
    \caption{Von Neumann entropy between the electronic and nuclear subspaces (Sec.~\ref{sec:qi}) at three points along the \ac{h2} dissociation curve.
    The entropy increases with the extension of the bond length and then slightly declines upon dissociation. Increasing the order of excitations in the quantum computing \ac{neo}\ac{ucc} ansatz, one can accurately reproduce the reference value obtained with the classical approach, \ac{neofci}.}
    \centering
    \begin{tabular}{@{}llrrr@{}}
        \toprule
         &                                                      & \multicolumn{3}{c}{Separation}\\
        \cmidrule{3-5}
         & Method               & 0.7414~{\AA}   & 1.9582~{\AA} & 3.1751~{\AA} \\
        \midrule
         &\ac{neofci}           & 0.0069 & 0.0103 & 0.0097 \\
        \midrule
        \multirow{3}{*}{\rotatebox[origin=c]{90}{\scriptsize\ac{neo}\ac{ucc}}}
         &SD                    & 0.0062 & 0.0050 & 0.0037 \\
         &SDT                   & 0.0068 & 0.0104 & 0.0098\\
         &SDTQ                  & 0.0068 & 0.0104 & 0.0098 \\
         \\[-0.3cm]
        \bottomrule
        \end{tabular}
    \label{tab:h2_von_neumann}
\end{table}

Although a \ac{ucc} type ansatz can be very accurate, in general, they cannot yet be efficiently utilized on current quantum 
hardware. Thus we present an optimal strategy for the application of the 
TwoLocal ansatz with \ac{neo} Hamiltonian employing parameter initialization procedures discussed in Section~\ref{sec:partran}. Similar to the \ac{neo}\ac{ucc} case we 
first optimized the TwoLocal ansatz for the electronic Hamiltonian and relaxed electronic 
parameters further, employing the $P_e\hat{H}_{\text{\ac{neo}}}P_e$ Hamiltonian. For both Hamiltonians 
we used TwoLocal ansatz with 8 entangling layers, see Figure~\ref{fig:TwoLocalGeneral} 
and Figure~\ref{circ:TLex}. As can be seen from Table~\ref{tab:h2TLbo} the TwoLocal ansatz performs similarly to the \acs{neoucc} ansatz both in terms of accuracy and the validity of the advanced parameter initialization. However, the 8-qubit \ac{neo} Hamiltonian represents a difficult optimization case for the TwoLocal ansatz with 8 entangling layers if ordinary parameter initialization is used.
Initial guess parameters are usually set randomly which often can not even deliver optimized energy below  the \ac{neohf} energy. One can see from Table~\ref{tab:h2TLneo} that for the best optimization case, the error is still larger than for 5-qubit TwoLocal ansatz applied on $P_e\hat{H}_{\text{\ac{neo}}}P_e$. 

If one sets the initial guess to the relaxed electronic parameters in TwoLocal ansatz according to the ``expanded'' variant with 8 entangling layers, see Figures~\ref{fig:TwoLocalGeneral} and \ref{circ:TLex}, the energy decreases slightly below the ground state energy for the $P_e\hat{H}_{\text{\ac{neo}}}P_e$ Hamiltonian. Although the ``expanded'' variant of the electronic parameter initialization delivers lower energy than ordinary optimization the error amounts to $10^{-3}$ Hartree. Expanding the relaxed electronic 5-qubit TwoLocal ansatz  with 3 nuclear qubits we continue stacking entangling layers between electronic and nuclear qubits only according to Section~\ref{sec:partran}. This ``stacked'' variant of electronic parameter initialization, Figure~\ref{circ:TLst}, performs better, see Table~\ref{tab:h2TLneo}, giving an error of about $2\times10^{-4}$ Hartree when 14 stacked layers are used. Although the proper parameter initialization requires an even number of stacked electronic-nuclear entangling layers the cases with an odd number perform equally well. Initial energy in \ac{vqe} optimization is above \ac{neohf} energy if an odd number of entangling layers is used whereas, for an even number, it is exactly the ground state energy for $P_e\hat{H}_{\text{\ac{neo}}}P_e$. Obviously, it did not affect the \ac{vqe} optimization using the \ac{cg} and \ac{slsqp}. It should be also noted that \ac{cg} optimizer performed best in the case of TwoLocal ansatz, while \ac{slsqp} and \ac{cobyla} deliver similar results.

\begin{table}[htb]
\caption{\ac{h2} ground state energies calculated with the TwoLocal and two classical approaches: \ac{fci} and ``exact diagonalization''. Two Hamiltonians are used: the electronic Hamiltonian ($\hat{H}_{\text{El.}}$) and the projecetd \ac{neo} Hamiltonian ($P_e\hat{H}_{\text{\ac{neo}}}P_e$). 
The TwoLocal electronic parameters reproducing the \ac{fci} wave function provide a good initial guess for the \ac{neo} calculation with the Hamiltonian $P_e\hat{H}_{\text{\ac{neo}}}P_e$, leading to an error of $8.35 \times 10^{-5}$~Ha. Further optimization of parameters (``relaxed'') reduces the error to $3.83 \times 10^{-7}$~Ha compared to the ground state energy of the $P_e\hat{H}_{\text{\ac{neo}}}P_e$ (``exact diagonalization'').
}
\label{tab:h2TLbo}
\begin{center}
\begin{tabular}{llcr}
\toprule
Hamiltonian & Method & Parameters & Energy/Ha \\
\midrule
$\hat{H}_{\text{El.}}$ & \acs{fci} &            & -1.151683\\ 
            & TwoLocal  & electronic            & -1.151683\\
\midrule
$P_e\hat{H}_{\text{\ac{neo}}}P_e$ & exact diagonalization & & -1.065040\\
                  & TwoLocal  & electronic      & -1.064956\\
               & TwoLocal  & relaxed & -1.065039\\
\bottomrule
\end{tabular}
\end{center}
\end{table}

\begin{table}[htb]
\caption{Ground state energies for the \acs{h2} \acs{neo} Hamiltonian obtained with various implementations of the TwoLocal quantum ansatz and with the classical \ac{neofci} approach. The ``advanced'' initialization scheme for electronic parameters performs significantly better than the ``ordinary'' one for both \ac{cg} and \ac{slsqp} optimizers. The ``stacked'' initialization with 14 entangling layers delivers the most accurate result with an error of $2\times10^{-4}$~Ha using the \ac{cg} optimizer.}
\label{tab:h2TLneo}
\begin{center}
\begin{tabular}{@{}ll@{\hskip 5mm}r@{\hskip 5mm}rr@{}}
\toprule
& Method & Layers & \multicolumn{2}{c}{Energy/Ha} \\
\midrule
& &  & \ac{slsqp}         & \ac{cg} \\
\midrule
\multirow{10}{*}{\rotatebox[origin=c]{90}{TwoLocal}} & Ordinary & & -1.013923 & -1.014310\\
\cmidrule{2-5}
& Expanded & & -1.065054 & -1.065172\\
\cmidrule{2-5}
 & Stacked & $d = 3$  &  -1.065175 & -1.065633\\
 & & $d = 4$  & -1.065056 & -1.065381\\
 & & $d = 10$ & -1.065055 & -1.065663\\
 & & $d = 11$ & -1.064975 & -1.065814\\
 & & $d = 12$ & -1.065055 & -1.065720\\
 & & $d = 13$ & -1.064939 & -1.065691\\
 & & $d = 14$ & -1.065054 & -1.065889\\
 & & $d = 20$ & -1.065056 & -1.065666\\
 \midrule
& \ac{neofci} &  & \multicolumn{2}{c}{-1.066121}\\
\bottomrule
\end{tabular}
\end{center}
\end{table}

\subsection{Malonaldehyde}\label{sec:malon}
We start our malonaldehyde investigation with the ``2-center $C_{2v}$'' setup, which corresponds to the situation in which the proton is equally shared between the two oxygen atoms (see Figure~\ref{fig:malon_BO_PES}).  
A similar setup was already proposed in Ref.~\cite{webb2002} for the evaluation of double well splitting. However, our aim is different and consists mainly in providing benchmark results (more specifically, ground state energies) for the validation of the quantum \acs{neo} algorithm. 
Using the same transformations as in Section~\ref{sec:h2}, we converted the second quantized Hamiltonian based on 16 spin orbitals to an 8-qubit Hamiltonian, 5 qubits spanning the electronic subspace, and 3 qubits spanning the nuclear one.
Increasing excitation order from double to quadruple in the \acs{neoucc} ansatz, we could recover the \ac{neocasci} reference results and reach $10^{-6}$ Hartree accuracy, as shown in Table~\ref{tab:Malon}. The results for the entanglement entropy are presented in Table~\ref{tab:Malon} and are also in very good agreement with the reference values. 

For evaluation of the tunneling barrier, we used the ``3-center $C_{2v}$'' (top of the barrier) and ``1-center $C_{s}$'' (bottom of the potential, see Figure~\ref{fig:malon_BO_PES}) setups as described in Section~\ref{sec:systems}. Based on the \ac{neocasci} calculations at these two setups, the barrier is estimated to be 0.005011 Hartree. This is rather close to the values obtained in Ref.~\cite{wang2008a} (full dimensional calculations with near basis-set-limit frozen-core CCSD(T)) and Ref.~\cite{list2020} 
({\it in silico} transient X-ray absorption spectroscopy) of 
0.0065 Hartree and 0.0061 Hartree, respectively. 
On the other hand, using a fully classical approach such as \ac{casci} for the evaluation of the barrier, we observed a severe overestimation of the barrier, which reaches a value of 0.01196 Hartree (see Figure~\ref{fig:malon_BO_PES} and table~\ref{tab:Malon}). 
Note that in Figure~\ref{fig:malon_BO_PES} we also report the full potential energy surface computed with \ac{mp2} and a classical proton. However, at this small basis set size for the expansion of the electronic wave function, we can only expect a qualitative description of the process, which we only report as guidance for the reader's eyes. 
To account for nuclear quantum effects, we repeat the same calculation using our proposed \acs{neoucc} algorithm. 
In this setup, the qubit Hamiltonian for the ``3-center $C_{2V}$'' setup is spanned by 10 qubits, 5 qubits for electronic subspace and 5 qubits for nuclear subspace. 
The ``1-center $C_{S}$'' qubit Hamiltonian is instead spanned by 7 qubits, 6 qubits for electronic subspace and 1 qubit for nuclear subspace. 
The energies for both systems estimated with various \acs{neoucc} schemes are presented in Table~\ref{tab:Malon}. 
Already with a rather modest \acs{neouccsd} ansatz, we observe very good agreement with the \ac{neocasci} reference, with a deviation smaller than $10^{-4}$ Hartree. 
The error decreases upon inclusion of higher order excitation operators in \acs{neoucc} and gets below $2\times10^{-6}$ Hartree for \acs{neouccsd}TQ. 
These calculations confirm the quality and the potential of the \acs{neoucc} ansatz for the calculation of quantum nuclei effects and prove its fast convergence towards the exact solution.

In Table~\ref{tab:Malon} we summarize the values for the barrier energies together with the ones for the von Neumann entropy, $s_n$ (Eq.~\eqref{eq:vonNeuman}), associated with the entanglement between the nuclear and electronic subsystems. 
Also in this case the NEOUCC results are in very good agreement with the \ac{neocasci} references for all settings: ``1-center $C_{s}$'', ``2-center $C_{2v}$'', and ``3-center $C_{2v}$''.
Assuming that the estimate for the $C_{s}$ system is not very sensitive to the basis set size, our calculations support the picture in which the level of electron-nuclear entanglement increases as the proton is transferred to the top of the barrier (however, this observation can not be conclusive in view of the small basis set used for the electronic and nuclear problem, as well as of the rigidity imposed to the molecular scaffold). 
Further investigations are needed to shed light on the convergence of the entanglement entropy between electronic and nuclear subsystems as a function of the nuclear basis set size.

\begin{table}[htb]
\caption{Energies and von Neumann entropies, $s_n$ (Eq.~\eqref{eq:vonNeuman}), of malonaldehyde obtained with different \acs{neoucc} parameterizations and \acs{neo} Hamiltonians based on ``3-center $C_{2v}$'', ``2-center $C_{2v}$'', and ``1-center $C_{s}$'' setups. The barrier height, $\Delta E$, is evaluated based on the difference between energies corresponding to ``3-center $C_{2v}$'' and ``1-center $C_{s}$'' setups. We observe good agreement with barrier height obtained in Ref.~\cite{wang2008a} and Ref.~\cite{list2020} for all variants of \acs{neoucc}, while the electronic \acs{casci} method severely overestimates it.}
\label{tab:Malon}
\begin{center}
\begin{tabular}{@{}llrrrrrrr@{}}
\toprule
 &            &  \multicolumn{2}{c}{``1-center $C_{s}$''}   &\multicolumn{2}{c}{``2-center $C_{2v}$''}&                \multicolumn{2}{c}{``3-center $C_{2v}$''} & \\
\midrule
 & Method              & Energy/Ha & Entropy& Energy/Ha & Entropy & Energy/Ha & Entropy& $\Delta$ E/Ha \\
\midrule
&Ref.~\cite{wang2008a} &             &        &             &         &             &        & 0.006534 \\
&Ref.~\cite{list2020} &             &        &             &         &             &        & 0.006056 \\
& \ac{casci}           & -265.528820 &        &             &         & -265.516859 &        & 0.011962 \\
 &\ac{neocasci}        & -265.490948 & 0.0000 & -265.452062 & 0.0866  & -265.485937 & 0.0044 & 0.005011 \\
\midrule
\\
 &SD                   & -265.490685 & 0.0000 & -265.451123 & 0.0616 & -265.485761 & 0.0031 & 0.004924 \\
 &SDT                  & -265.490943 & 0.0000 & -265.452051 & 0.0862 & -265.485936 & 0.0044 & 0.005007 \\
 \multirow{-4}{*}{\rotatebox[origin=c]{90}{\ac{neo}\ac{ucc}}}
 &SDTQ                 & -265.490947 & 0.0000 & -265.452062 & 0.0866 & -265.485939 & 0.0045 & 0.005009 \\
\midrule
\end{tabular}
\end{center}
\end{table}
\section{Discussion and Conclusions}\label{sec:conclusion}

In this work, we developed and demonstrated a quantum algorithm for the  treatment of quantum nuclear degrees of freedom on near-term quantum computers, based on the \acs{neo} approach~\cite{webb2002}. 
This approach allows the description of both components of molecular systems, \textit{i.e.}, electrons and nuclei, at the same footing, namely at a quantum level, enabling the correct description of important nuclear quantum effects such as proton tunneling and non-adiabatic effects. 
From the quantum computing perspective, a quantum treatment of the full molecular wave function gives us the opportunity to extend the potential exponential quantum advantage already investigated for the electronic subsystem to the multi-component electron-nuclear wave function. 
In addition, this method will enable a rigorous estimation of the level of entanglement between the electronic and nuclear subsystems, providing a new and interesting tool for the understanding of post-\ac{bo} effects. 
In order to make our approach more suited to near-term quantum computing, we implemented an embedding scheme, which allows the restriction of the nuclear quantum wave function to a selected number of atoms, while keeping the rest of the nuclei described at a classical level (\textit{i.e.}, as point charges).  

The ability of \ac{neo}\ac{ucc} ans\"atze to deliver the ground state energy in noiseless \ac{vqe} calculation was proven in two main applications: the hydrogen molecule, and malonaldehyde. In the latter case, only the shared proton between the two carbonyl groups was ``quantized'' while all other nuclei in the molecules were treated classically.  
Our results agree with conventional \ac{neofci} and \ac{neocasci} calculations to within an error of about  $10^{-6}$ Hartree.  

Our \acs{neoucc} calculations revealed that apart from the high contribution to the correlation energy from interactions between electrons, a substantial part of correlation energy is attributed to electron-nuclear interaction. 
Specifically, double electron-nuclear excitations are required for reducing the error below $10^{-4}$ Hartree, and only after the addition of triplet T$^{(2,1)}$ excitations, one can reach an accuracy of about $10^{-6}$ Hartree. 
The importance of nuclear-electron quantum correlations is also demonstrated by the significant amount of nuclear-electron von Neumann entropy, which increases at bond breaking. 

Moreover, we investigated the possibility of utilizing hardware-efficient ans\"atze as more compact representations of the \ac{neo} wave functions. We showed that the parameter initialization scheme plays an important role in the optimization process. 
Namely, to reach an accuracy below $5\times10^{-4}$ Hartree, we introduced several ways of initialization of the \ac{neo}  electronic wave function parameters from the converged electronic calculation. 
Such an efficient initialization procedure is not restricted to the use of hardware-efficient ans\"atze, and can also significantly increase convergence in the case of \acs{neoucc} wave functions. 
In addition, the same scheme can also be employed beyond the typical cases for which the \ac{bo} separation is employed, for instance in the case of weakly interacting molecular fragments.

Furthermore, we introduced a scheme for reducing the dimensionality of the effective Hamiltonian through the exploitation of those symmetries inherent in the \acs{neo} approach. 
Specifically, we have generalized the two-qubit reduction, already employed for the electronic wave function, extending it to the nuclear subsystem. The dimensionality of the nuclear Hilbert space is further reduced by leveraging the point group symmetry inherent in the molecular system. 
As this last step might have undesired consequences on the initialization of the electronic parameter, we have introduced and discussed a procedure for resolving the issue. In the last step, when dealing with nuclei of the same spin polarization 
we can additionally project out unpopulated spin orbitals from the Hamiltonian and remove corresponding qubits. 
This was demonstrated in this work for both orthohydrogen spin isomer and malonaldehyde with a single proton described as a quantum particle. 
Thus, for the qubit Hamiltonian in the case of orthohydrogen, we succeeded in reducing 
the number of qubits from 16 to 8. Similarly, for the malonaldehyde instead of using a 20-qubit Hamiltonian, we  demonstrated the possibility to safely reduce the system to an effective 10-qubit Hamiltonian. 
Reduction in the number of qubits leads to a significant decrease in the number of two-qubit entangling gates in the ansatz as well as the total number of terms in Hamiltonian required for measurement in \ac{vqe}, making the entire algorithm more suited for near-term hardware calculations.

In conclusion,
we have proposed and implemented  a resource-efficient quantum algorithm for the simulation of molecular systems, where both electrons and nuclei are treated quantum mechanically within the \ac{neo} approach.
Although the execution of the proposed \acs{neo}/\ac{vqe} algorithm on state-of-the-art quantum hardware for systems of the size of malonaldehyde (with one quantum hydrogen atom) and orthohydrogen (with two quantum hydrogen atoms) is still premature, the developed machinery allowed us to benchmark our approach using \ac{vqe} state vector simulations. 
By means of our applications, we could show the reliability and versatility of the \acs{neoucc} ansatz, which is able of capturing nuclear-electron wave function energies and subsystem entropies with controlled accuracy.
In particular, we demonstrated the possibility to evaluate the tunneling barrier for malonaldehyde with increasing accuracy by incorporating quantum nuclear effects into the calculation with the proposed \acs{neoucc} algorithm.
Building on these very promising initial results, we aim in the future to further generalize this framework by including sampling techniques and the possibility of investigating quantum electron-proton dynamics, opening up --- thanks to the favorable scaling of the proposed quantum algorithm ($\mathcal{O}(N_{b}^{e}+N_{b}^{n})^4$, where $N_{b}^{e}$ and $N_{b}^{n}$ are the numbers of basis set functions for electrons and nuclei respectively) --- new avenues in the study of nuclear quantum effects in catalysis.

\section{Acknowledgements}
This work was supported by the Hartree National Centre for Digital Innovation, a collaboration between STFC and IBM. The authors thank Lars Tornberg for the helpful discussions. This research was supported by funding from Horizon 2020 via NEASQC project (grant number 951821),
Wallenberg Center for Quantum Technology (WACQT), and from the NCCR MARVEL, a National Centre of Competence in Research, funded by the Swiss National Science Foundation (grant number 205602). IBM, the IBM logo, and ibm.com are trademarks of International Business Machines Corp., registered in many jurisdictions worldwide. Other product and service names might be trademarks of IBM or other companies. The current list of IBM trademarks is available at \url{https://www.ibm.com/legal/copytrade}.

\appendix
\section{Single orbital entropy and quantum information}

Single orbital entropies~\cite{lege2006,bogu2012b} are calculated as
\begin{equation}
    so_i = -\sum_{\alpha=1}^4 \omega_{i,\alpha} \ln (\omega_{i,\alpha}),
\end{equation}
where $\omega_{i,\alpha}$ are the eigenvalues of the one-orbital reduced density matrix~\cite{bogu2015,bogu2012b}, the index $i$ runs over molecular orbitals, and $\alpha$ runs over the four possible occupations: $--,-\uparrow,\downarrow-,\downarrow\uparrow$. The sum over all single orbital
entropies, the total quantum information~\cite{bogu2012b}, is given by
\begin{equation}
QI = \sum_{i}so_i \, .
\end{equation} 
The results for \ac{h2} (studied in Section~\ref{sec:h2})
and malonaldehyde in ``2-center $C_{2v}$'' setting (studied in Section~\ref{sec:malon}) are  shown in Tables~\ref{tab:ucch2QI} and~\ref{tab:uccQI}.
\begin{table}[htb]
\caption{The single orbital entropies, $so_i$, and quantum information (total, electronic, and nuclear) for \acs{h2} (studied in Section~\ref{sec:h2})
evaluated with \ac{neo}\ac{ucc}SDT.}
\label{tab:ucch2QI}
\begin{center}
\begin{tabular}{llr}
 & orbital    & $so_i$ \\
\hline
 &1 & 0.0793 \\
 &2 & 0.0630 \\
 &3 & 0.0221 \\
 \multirow{-4}{*}{\rotatebox[origin=c]{90}{electronic}}
 &4 & 0.0397 \\
 \hline
 &$QI(\text{electronic})$ & 0.2041\\
 \hline
 &5 & 0.0035 \\
 &6 & 0.0035 \\
 &7 & 0.0035 \\
 \multirow{-4}{*}{\rotatebox[origin=c]{90}{nuclear}}
 &8 & 0.0035 \\
 \hline
 &$QI(\text{nuclear})$ & ($\approx$ 6.4\%) 0.0139\\
 \hline
 &$QI(\text{total})$   & 0.2179\\
\end{tabular}
\end{center}
\end{table}

\begin{table}[htb]
\caption{The single orbital entropies, $so_i$, and quantum information (total, electronic, and nuclear) for malonaldehyde (``2-center $C_{2v}$'' setup)  evaluated with \ac{neo}\ac{ucc}SDTQ.}\label{tab:uccQI}
\begin{center}
\begin{tabular}{llr}
 &orbital    & $so_i$ \\
\hline
 &1 & 0.0896 \\
 &2 & 0.1955 \\
 &3 & 0.2049 \\
 \multirow{-4}{*}{\rotatebox[origin=c]{90}{electronic}}
 &4 & 0.0806 \\
 \hline
 &$QI(\text{electronic})$ & 0.5705\\
 \hline
 &5 & 0.0867 \\
 &6 & 0.0866 \\
 &7 & 0.0000 \\
 \multirow{-4}{*}{\rotatebox[origin=c]{90}{nuclear}}
 &8 & 0.0001 \\
 \hline
&$QI(\text{nuclear})$ & ($\approx$ 23\%) 0.1734\\
 \hline
& $QI(\text{total})$   & 0.7440\\
\end{tabular}
\end{center}
\end{table}

\section{Parameter initialization and convergence}\label{sec:PTconvergence}

\begin{figure}[htb]
    \centering
    \begin{subfigure}[b]{0.49\textwidth}
    \centering
    \includegraphics[width=\textwidth]{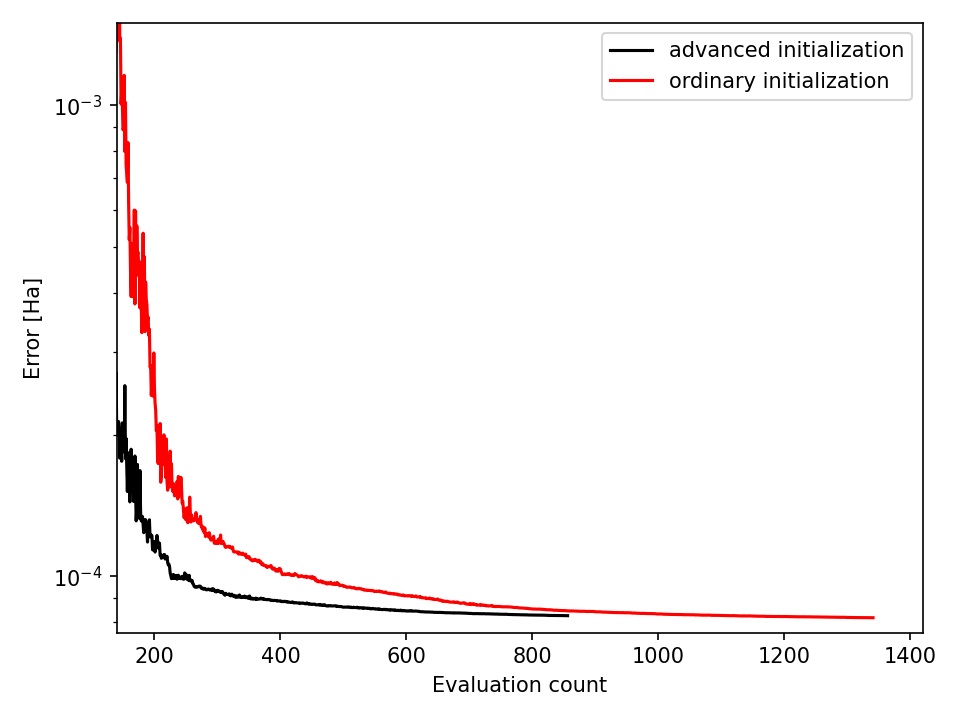}
    \caption{\acs{neouccsd}}
    \label{fig:h2cobyla}
    \end{subfigure}
    \begin{subfigure}[b]{0.49\textwidth}
    \centering
    \includegraphics[width=\textwidth]{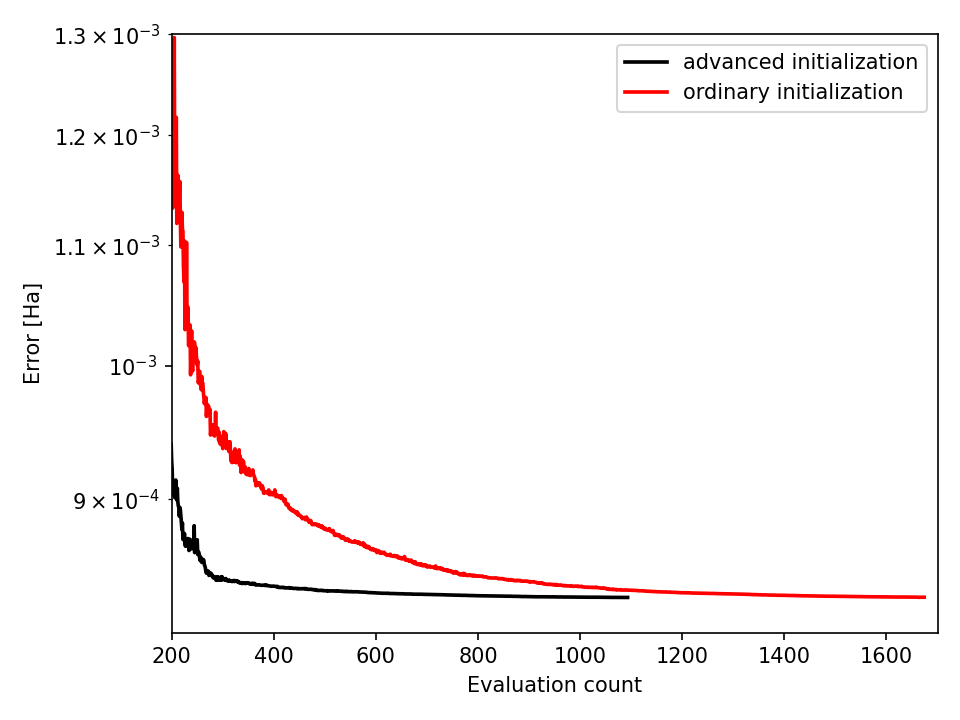}
    \caption{\acs{neouccsd}T$^{(1,2)}$}
    \label{fig:h2dsdcobyla}
    \end{subfigure}
    \\ 
    \begin{subfigure}[b]{0.49\textwidth}
    \centering
    \includegraphics[width=\textwidth]{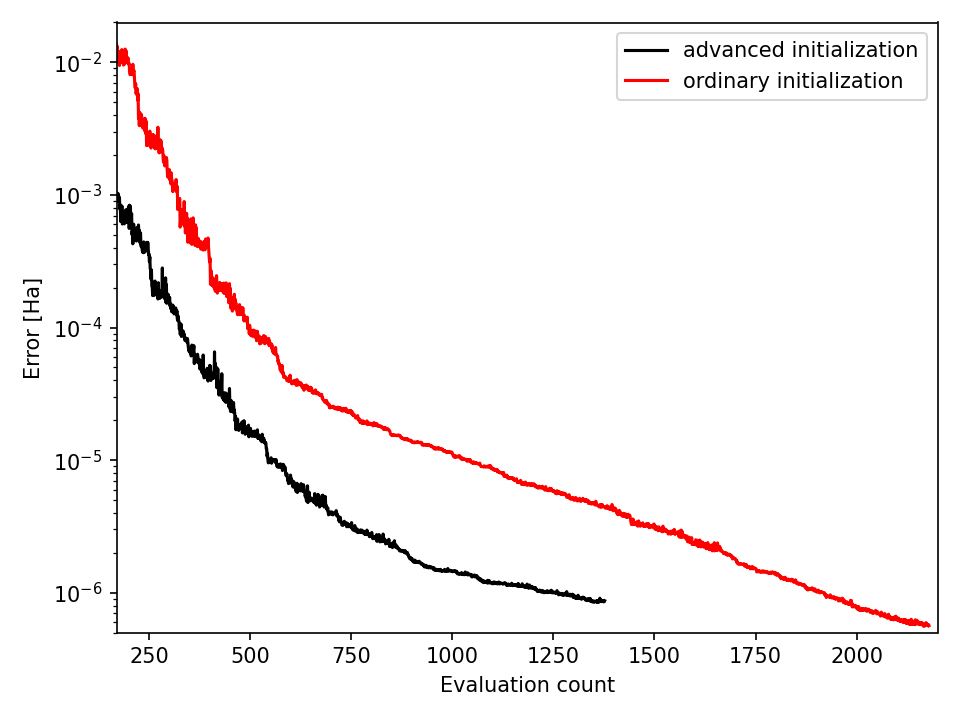}
    \caption{\acs{neouccsd}T$^{(2,1)}$}
    \label{fig:h2ddscobyla}
    \end{subfigure}
    \begin{subfigure}[b]{0.49\textwidth}
    \centering
    \includegraphics[width=\textwidth]{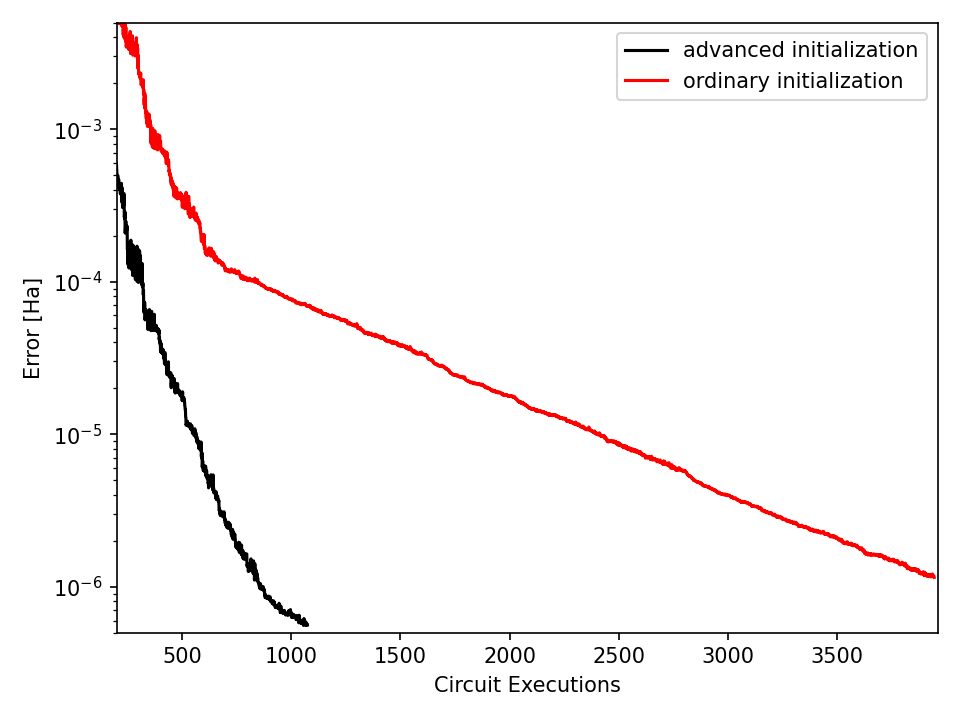}
    \caption{\acs{neouccsd}T}
    \label{fig:h2dtcobyla}
    \end{subfigure}
    \caption{Energy convergence plots for the H$_2$ molecule with different \acs{neoucc} ans\"atze and two initialization schemes. The classical optimization of the circuit parameters is done employing the \ac{cobyla} optimizer. The ``advanced'' initialization scheme introduced in Section~\ref{sec:ansatzinit} (red lines) clearly outperforms the ``ordinary'' initialization procedure (black lines) for what concerns the speed of convergence, especially for large number of parameters (with SDT excitations).}
    \label{fig:h2dconv}
\end{figure}

Figure~\ref{fig:h2dconv} shows the different convergence profiles for the optimizations using ``advanced'' and ``ordinary'' initialization schemes with the \acs{neoucc} ans\"atze and the \ac{cobyla} optimizer. 
As expected, we notice a faster convergence when the parameters precomputed for the electronic subsystem are transferred to the \acs{neoucc} ansatz 
 (Figure~\ref{fig:h2dconv}). 
 Also, the initial energy values for the ``advanced'' initialization are lower than the ones using the ``ordinary'' initialization. 
In addition, our calculations show that as the complexity (and therefore the accuracy) of the ansatz increases, the effect of ``advanced'' initialization of the parameters becomes more important.
Specifically, the convergence rate decreases upon the addition of triple excitations in going from \acs{neouccsd}, Figure~\ref{fig:h2cobyla}, to \acs{neouccsd}T$^{(2,1)}$ and \acs{neouccsd}T$^{(1,2)}$, Figures~\ref{fig:h2ddscobyla} and~\ref{fig:h2dsdcobyla}, respectively. 
Even though the number of electronic parameters is the same in both initialization procedures, the benefit of the `advanced' procedure is indisputable, especially for the \acs{neouccsd}T ansatz where the convergence rate is increased more than threefold.

\section{Resource estimation}\label{sec:resource_estimate}
Both the number of qubits and the depth of the quantum circuit play a crucial role in determining the performance of quantum algorithms in near-term quantum computers. 
It is therefore important to analyze the hardware requirements needed to implement our quantum algorithm. 
As for the quantum gates, the number of entangling gates can strongly impact the circuit noise. 
In Table~\ref{tab:resources} we summarize the number of qubits, two-qubit entangling gates, and terms in the qubit Hamiltonian needed for the implementation of a \ac{vqe} calculation for \ac{h2}, as reported in Section~\ref{sec:h2}. 
Specifically, these numbers refer to the final counts obtained after applying three different reduction schemes sequentially: (i) four-qubit tapering, (ii) point group (PG) symmetry, and (iii) projection onto a selected spin state. 
As wave function ans\"atze we employ TwoLocal (8 layers for the electronic part and 14 nuclear-electronic entangling layers) and \acs{neouccsd}, since they showed similar accuracy (see Section~\ref{sec:h2}). 
With the proposed schemes, we achieved a significant reduction --- compared to the standard procedure --- in the number of qubits and  Hamiltonian terms, as well as in the number of entangling gates, for both ans\"atze.

\begin{table}[htb]
    \centering
    \caption{Estimated resource requirements for the \ac{h2} molecule (see Section~\ref{sec:h2}) using various reduction schemes. Shown are the number of qubits, two-qubit gates in TwoLocal as well as in \acs{neouccsd} ans\"atze, and the number of Pauli terms ($|\{H\}|$) in the encoded Hamiltonian. All values are calculated using parity encoding. For the TwoLocal ansatz, 8 electron-to-electron entangling layers are applied to the electronic subsystem together with 14 electron-to-proton entangling layers.}
    \begin{tabular}{@{}r@{\hskip 2mm}rrrr@{}}
        \toprule
         & & \multicolumn{2}{c}{\# Two-qubit Gates} & \\
         \cmidrule{3-4}
        Reduction & \# Qubits & TwoLocal & \acs{neouccsd} & $|\{H\}|$ \\
        \midrule
        None                        & 16 & 968 & 2546 & 861 \\
        +Four-qubit                  & 12 & 558 & 2204 & 828 \\
        +PG symmetry                 & 11 & 465 & 1472 & 825 \\
        +Spin projection             &  8 & 255 & 1202 & 428 \\
        \bottomrule
    \end{tabular}
    \label{tab:resources}
\end{table}

\section{\ac{neo} qubit Hamiltonian for \ac{h2} in minimal basis set}\label{sec:minimalH2}

\begin{alignat}{6}\label{eq:minimalH2}
\hat H_{\text{\ac{neo}}} & =1.697878 \times IIII && -0.373036 \times ZIII && +0.356877 \times IZII && -0.373036 \times ZZII && -0.337003 \times IIZI\nonumber\\
                         & +1.105641 \times ZIZI && -0.337002 \times IZZI && +1.094360 \times ZZZI && -0.416401 \times IIIZ && +0.004053 \times ZIIZ\nonumber\\
                         & +0.004053 \times ZZIZ && -0.004053 \times IIZZ && -0.004053 \times IZZZ && +0.416401 \times ZZZZ && +0.174434 \times XIII\nonumber\\
                         & -0.174434 \times XZII && -0.006416 \times IXII && +0.006416 \times ZXZI && -0.004740 \times YYII && -0.004740 \times XXZI\nonumber\\
                         & +0.004740 \times YYIZ && +0.004740 \times XXZZ && -0.186061 \times IIXI && +0.186061 \times IZXI && -0.748764 \times XIXI\nonumber\\
                         & +0.748764 \times XZXI && +0.004740 \times ZXXI && +0.004740 \times IYYI && -0.004740 \times ZXXZ && -0.004740 \times IYYZ\nonumber\\
                         & -0.006416 \times XXXI && -0.006416 \times YXYI && +0.181576 \times IIIX && +0.174434 \times XIIX && -0.174434 \times XZIX\nonumber\\
                         & -0.006416 \times IXIX && +0.006416 \times ZXZX && -0.186061 \times IIXX && +0.186061 \times IZXX && -0.006416 \times XXXX\nonumber\\
                         & -0.006416 \times YXYX &&&&&
\end{alignat}

\begin{thebibliography}{10}

\bibitem{markland2018nuclear}
Thomas~E Markland and Michele Ceriotti.
\newblock Nuclear quantum effects enter the mainstream.
\newblock {\em Nature Reviews Chemistry}, 2(3):1--14, 2018.

\bibitem{pereyaslavets2018importance}
Leonid Pereyaslavets, Igor Kurnikov, Ganesh Kamath, Oleg Butin, Alexey
  Illarionov, Igor Leontyev, Michael Olevanov, Michael Levitt, Roger~D
  Kornberg, and Boris Fain.
\newblock On the importance of accounting for nuclear quantum effects in ab
  initio calibrated force fields in biological simulations.
\newblock {\em Proceedings of the National Academy of Sciences},
  115(36):8878--8882, 2018.

\bibitem{heyes2009nuclear}
Derren~J Heyes, Michiyo Sakuma, Sam~P de~Visser, and Nigel~S Scrutton.
\newblock Nuclear quantum tunneling in the light-activated enzyme
  protochlorophyllide oxidoreductase.
\newblock {\em Journal of Biological Chemistry}, 284(6):3762--3767, 2009.

\bibitem{rossi2016anharmonic}
Mariana Rossi, Piero Gasparotto, and Michele Ceriotti.
\newblock Anharmonic and quantum fluctuations in molecular crystals: A
  first-principles study of the stability of paracetamol.
\newblock {\em Physical Review Letters}, 117(11):115702, 2016.

\bibitem{vardi2015nuclear}
Alexandra Vardi-Kilshtain, Neta Nitoker, and Dan~Thomas Major.
\newblock Nuclear quantum effects and kinetic isotope effects in enzyme
  reactions.
\newblock {\em Archives of Biochemistry and Biophysics}, 582:18--27, 2015.

\bibitem{raugei2003nuclear}
Simone Raugei and Michael~L Klein.
\newblock Nuclear quantum effects and hydrogen bonding in liquids.
\newblock {\em Journal of the American Chemical Society}, 125(30):8992--8993,
  2003.

\bibitem{kim2017temperature}
Kyung~Hwan Kim, Harshad Pathak, Alexander Sp{\"a}h, Fivos Perakis, Daniel
  Mariedahl, Jonas~A Sellberg, Tetsuo Katayama, Yoshihisa Harada, Hirohito
  Ogasawara, Lars~GM Pettersson, et~al.
\newblock Temperature-independent nuclear quantum effects on the structure of
  water.
\newblock {\em Physical Review Letters}, 119(7):075502, 2017.

\bibitem{pamuk2012anomalous}
Bet Pamuk, Jose~M Soler, R~Ram{\'\i}rez, CP~Herrero, PW~Stephens, PB~Allen, and
  M-V Fern{\'a}ndez-Serra.
\newblock Anomalous nuclear quantum effects in ice.
\newblock {\em Physical Review Letters}, 108(19):193003, 2012.

\bibitem{webb2002}
Hammes-Schiffer~{S.} Webb~{S. P.}, Iordanov~{T.}
\newblock Multiconfigurational nuclear-electronic orbital approach:
  Incorporation of nuclear quantum effects in electronic structure
  calculations.
\newblock {\em J. Chem. Phys.}, 117:4106--4118, 2002.

\bibitem{simm2013}
Benjamin Simmen, Edit M{\'{a}}tyus, and Markus Reiher.
\newblock {Elimination of the translational kinetic energy contamination in
  pre-Born–Oppenheimer calculations}.
\newblock {\em Molecular Physics}, 111(14-15):2086--2092, 2013.

\bibitem{naka2002}
Hiromi Nakai.
\newblock Simultaneous determination of nuclear and electronic wave functions
  without born–oppenheimer approximation: Ab initio no+mo/hf theory.
\newblock {\em International Journal of Quantum Chemistry}, 86(6):511--517,
  2002.

\bibitem{lee2022evidence}
Seunghoon Lee, Joonho Lee, Huanchen Zhai, Yu~Tong, Alexander~M. Dalzell,
  Ashutosh Kumar, Phillip Helms, Johnnie Gray, Zhi-Hao Cui, Wenyuan Liu,
  Michael Kastoryano, Ryan Babbush, John Preskill, David~R. Reichman, Earl~T.
  Campbell, Edward~F. Valeev, Lin Lin, and Garnet Kin-Lic Chan.
\newblock Is there evidence for exponential quantum advantage in quantum
  chemistry?
\newblock {\em arXiv:2208.02199}, 2022.

\bibitem{kjaergaard2020superconducting}
Morten Kjaergaard, Mollie~E Schwartz, Jochen Braum{\"u}ller, Philip Krantz,
  Joel I-J Wang, Simon Gustavsson, and William~D Oliver.
\newblock Superconducting qubits: Current state of play.
\newblock {\em Annual Review of Condensed Matter Physics}, 11:369--395, 2020.

\bibitem{huang2020superconducting}
He-Liang Huang, Dachao Wu, Daojin Fan, and Xiaobo Zhu.
\newblock Superconducting quantum computing: a review.
\newblock {\em Science China Information Sciences}, 63(8):1--32, 2020.

\bibitem{Somma2002}
R.~Somma, G.~Ortiz, J.~E. Gubernatis, E.~Knill, and R.~Laflamme.
\newblock Simulating physical phenomena by quantum networks.
\newblock {\em Phys. Rev. A}, 65:042323, Apr 2002.

\bibitem{wecker2015}
Dave Wecker, Matthew~B. Hastings, Nathan Wiebe, Bryan~K. Clark, Chetan Nayak,
  and Matthias Troyer.
\newblock Solving strongly correlated electron models on a quantum computer.
\newblock {\em Phys. Rev. A}, 92:062318, Dec 2015.

\bibitem{bauer2016hybrid}
Bela Bauer, Dave Wecker, Andrew~J. Millis, Matthew~B. Hastings, and Matthias
  Troyer.
\newblock Hybrid quantum-classical approach to correlated materials.
\newblock {\em Phys. Rev. X}, 6:031045, Sep 2016.

\bibitem{smith_simulating_2019}
Adam Smith, M.~S. Kim, Frank Pollmann, and Johannes Knolle.
\newblock Simulating quantum many-body dynamics on a current digital quantum
  computer.
\newblock {\em npj Quantum Information}, 5(1):106, November 2019.

\bibitem{cade2020}
Chris Cade, Lana Mineh, Ashley Montanaro, and Stasja Stanisic.
\newblock Strategies for solving the fermi-hubbard model on near-term quantum
  computers.
\newblock {\em Phys. Rev. B}, 102:235122, Dec 2020.

\bibitem{chiesa_quantum_2019}
A.~Chiesa, F.~Tacchino, M.~Grossi, P.~Santini, I.~Tavernelli, D.~Gerace, and
  S.~Carretta.
\newblock Quantum hardware simulating four-dimensional inelastic neutron
  scattering.
\newblock {\em Nature Physics}, 15(May 2019):455--459, March 2019.

\bibitem{RevTacchino2020}
F.~Tacchino, A.~Chiesa, S.~Carretta, and D.~Gerace.
\newblock Quantum computers as universal quantum simulators: State-of-the-art
  and perspectives.
\newblock {\em Advanced Quantum Technologies}, 3(3):1900052, 2020.

\bibitem{Suchsland_2021}
Philippe Suchsland, Panagiotis~Kl. Barkoutsos, Ivano Tavernelli, Mark~H.
  Fischer, and Titus Neupert.
\newblock Simulating a ring-like hubbard system with a quantum computer.
\newblock {\em arXiv:2104.06428}, 2021.

\bibitem{miessen2021quantum}
Alexander Miessen, Pauline~J. Ollitrault, and Ivano Tavernelli.
\newblock Quantum algorithms for quantum dynamics: A performance study on the
  spin-boson model.
\newblock {\em Phys. Rev. Research}, 3:043212, Dec 2021.

\bibitem{PhysRevResearch.4.043038}
Francesco Libbi, Jacopo Rizzo, Francesco Tacchino, Nicola Marzari, and Ivano
  Tavernelli.
\newblock Effective calculation of the green's function in the time domain on
  near-term quantum processors.
\newblock {\em Phys. Rev. Res.}, 4:043038, Oct 2022.

\bibitem{PhysRevResearch.4.043011}
Jacopo Rizzo, Francesco Libbi, Francesco Tacchino, Pauline~J. Ollitrault,
  Nicola Marzari, and Ivano Tavernelli.
\newblock One-particle green's functions from the quantum equation of motion
  algorithm.
\newblock {\em Phys. Rev. Res.}, 4:043011, Oct 2022.

\bibitem{martinez_real_time_2016}
E.~A. Martinez, C.~A. Muschik, P.~Schindler, D.~Nigg, A.~Erhard, M.~Heyl,
  P.~Hauke, M.~Dalmonte, T.~Monz, P.~Zoller, and R.~Blatt.
\newblock Real-time dynamics of lattice gauge theories with a few-qubit quantum
  computer.
\newblock {\em Nature}, 534(7608):516--519, June 2016.

\bibitem{klco_quantum-classical_2018}
N.~Klco, E.~F. Dumitrescu, A.~J. McCaskey, T.~D. Morris, R.~C. Pooser, M.~Sanz,
  E.~Solano, P.~Lougovski, and M.~J. Savage.
\newblock Quantum-classical computation of {Schwinger} model dynamics using
  quantum computers.
\newblock {\em Physical Review A}, 98(3):032331, September 2018.

\bibitem{roggero_dynamic_2019}
Alessandro Roggero and Joseph Carlson.
\newblock Dynamic linear response quantum algorithm.
\newblock {\em Phys. Rev. C}, 100:034610, Sep 2019.

\bibitem{mathis_toward_2020}
Simon~V. Mathis, Guglielmo Mazzola, and Ivano Tavernelli.
\newblock Toward scalable simulations of lattice gauge theories on quantum
  computers.
\newblock {\em Phys. Rev. D}, 102:094501, Nov 2020.

\bibitem{mazzola2021gauge}
Giulia Mazzola, Simon~V. Mathis, Guglielmo Mazzola, and Ivano Tavernelli.
\newblock Gauge-invariant quantum circuits for $u$(1) and yang-mills lattice
  gauge theories.
\newblock {\em Phys. Rev. Res.}, 3:043209, Dec 2021.

\bibitem{wu2021}
Sau~Lan Wu, Shaojun Sun, Wen Guan, Chen Zhou, Jay Chan, Chi~Lung Cheng, Tuan
  Pham, Yan Qian, Alex~Zeng Wang, Rui Zhang, Miron Livny, Jennifer Glick,
  Panagiotis~Kl. Barkoutsos, Stefan Woerner, Ivano Tavernelli, Federico
  Carminati, Alberto Di~Meglio, Andy C.~Y. Li, Joseph Lykken, Panagiotis
  Spentzouris, Samuel Yen-Chi Chen, Shinjae Yoo, and Tzu-Chieh Wei.
\newblock Application of quantum machine learning using the quantum kernel
  algorithm on high energy physics analysis at the lhc.
\newblock {\em Phys. Rev. Res.}, 3:033221, Sep 2021.

\bibitem{crippa2022}
Giuseppe Clemente, Arianna Crippa, and Karl Jansen.
\newblock Strategies for the determination of the running coupling of
  ($2+1$)-dimensional qed with quantum computing.
\newblock {\em Phys. Rev. D}, 106:114511, Dec 2022.

\bibitem{PhysRevResearch.3.033221}
Sau~Lan Wu, Shaojun Sun, Wen Guan, Chen Zhou, Jay Chan, Chi~Lung Cheng, Tuan
  Pham, Yan Qian, Alex~Zeng Wang, Rui Zhang, Miron Livny, Jennifer Glick,
  Panagiotis~Kl. Barkoutsos, Stefan Woerner, Ivano Tavernelli, Federico
  Carminati, Alberto Di~Meglio, Andy C.~Y. Li, Joseph Lykken, Panagiotis
  Spentzouris, Samuel Yen-Chi Chen, Shinjae Yoo, and Tzu-Chieh Wei.
\newblock Application of quantum machine learning using the quantum kernel
  algorithm on high energy physics analysis at the lhc.
\newblock {\em Phys. Rev. Res.}, 3:033221, Sep 2021.

\bibitem{schuhmacher2023unravelling}
Julian Schuhmacher, Laura Boggia, Vasilis Belis, Ema Puljak, Michele Grossi,
  Maurizio Pierini, Sofia Vallecorsa, Francesco Tacchino, Panagiotis
  Barkoutsos, and Ivano Tavernelli.
\newblock Unravelling physics beyond the standard model with classical and
  quantum anomaly detection.
\newblock {\em arXiv preprint arXiv:2301.10787}, 2023.

\bibitem{wozniak2023quantum}
Kinga~Anna Wo{\'z}niak, Vasilis Belis, Ema Puljak, Panagiotis Barkoutsos,
  G{\"u}nther Dissertori, Michele Grossi, Maurizio Pierini, Florentin Reiter,
  Ivano Tavernelli, and Sofia Vallecorsa.
\newblock Quantum anomaly detection in the latent space of proton collision
  events at the lhc.
\newblock {\em arXiv preprint arXiv:2301.10780}, 2023.

\bibitem{OMalley2016}
P.~J.~J. O'Malley, R.~Babbush, I.~D. Kivlichan, J.~Romero, J.~R. McClean,
  R.~Barends, J.~Kelly, P.~Roushan, A.~Tranter, and \emph{et al.}
\newblock Scalable quantum simulation of molecular energies.
\newblock {\em Phys. Rev. X}, 6:031007, Jul 2016.

\bibitem{kandala2017hardware}
Abhinav Kandala, Antonio Mezzacapo, Kristan Temme, Maika Takita, Markus Brink,
  Jerry~M Chow, and Jay~M Gambetta.
\newblock Hardware-efficient variational quantum eigensolver for small
  molecules and quantum magnets.
\newblock {\em nature}, 549(7671):242--246, 2017.

\bibitem{Barkoutsos2018}
Panagiotis~Kl Barkoutsos, Jerome~F. Gonthier, Igor Sokolov, Nikolaj Moll, Gian
  Salis, Andreas Fuhrer, Marc Ganzhorn, Daniel~J. Egger, Matthias Troyer,
  Antonio Mezzacapo, Stefan Filipp, and Ivano Tavernelli.
\newblock {Quantum algorithms for electronic structure calculations:
  Particle-hole Hamiltonian and optimized wave-function expansions}.
\newblock {\em Physical Review A}, 98(2):1--13, 2018.

\bibitem{cao2019quantum}
Yudong Cao, Jonathan Romero, Jonathan~P Olson, Matthias Degroote, Peter~D
  Johnson, M{\'a}ria Kieferov{\'a}, Ian~D Kivlichan, Tim Menke, Borja
  Peropadre, Nicolas~PD Sawaya, et~al.
\newblock Quantum chemistry in the age of quantum computing.
\newblock {\em Chemical Reviews}, 119(19):10856--10915, 2019.

\bibitem{McArdle2020}
Sam McArdle, Suguru Endo, Al\'an Aspuru-Guzik, Simon~C. Benjamin, and Xiao
  Yuan.
\newblock Quantum computational chemistry.
\newblock {\em Rev. Mod. Phys.}, 92:015003, Mar 2020.

\bibitem{sokolov2020}
Igor~O Sokolov, Panagiotis~Kl Barkoutsos, Pauline~J Ollitrault, Donny
  Greenberg, Julia Rice, Marco Pistoia, and Ivano Tavernelli.
\newblock Quantum orbital-optimized unitary coupled cluster methods in the
  strongly correlated regime: Can quantum algorithms outperform their classical
  equivalents?
\newblock {\em The Journal of Chemical Physics}, 152(12):124107, 2020.

\bibitem{ollitrault2019}
Pauline~J Ollitrault, Abhinav Kandala, Chun-Fu Chen, Panagiotis~Kl Barkoutsos,
  Antonio Mezzacapo, Marco Pistoia, Sarah Sheldon, Stefan Woerner, Jay~M
  Gambetta, and Ivano Tavernelli.
\newblock Quantum equation of motion for computing molecular excitation
  energies on a noisy quantum processor.
\newblock {\em Physical Review Research}, 2(4):043140, 2020.

\bibitem{sokolov_microcanonical_2021}
Igor~O. Sokolov, Panagiotis~Kl. Barkoutsos, Lukas Moeller, Philippe Suchsland,
  Guglielmo Mazzola, and Ivano Tavernelli.
\newblock Microcanonical and finite-temperature ab initio molecular dynamics
  simulations on quantum computers.
\newblock {\em Phys. Rev. Research}, 3:013125, Feb 2021.

\bibitem{kiss2022quantum}
Oriel Kiss, Francesco Tacchino, Sofia Vallecorsa, and Ivano Tavernelli.
\newblock Quantum neural networks force fields generation.
\newblock {\em Machine Learning: Science and Technology}, 3(3):035004, 2022.

\bibitem{barkoutsos2021quantum}
Panagiotis~Kl Barkoutsos, Fotios Gkritsis, Pauline~J Ollitrault, Igor~O
  Sokolov, Stefan Woerner, and Ivano Tavernelli.
\newblock Quantum algorithm for alchemical optimization in material design.
\newblock {\em Chemical science}, 12(12):4345--4352, 2021.

\bibitem{ollitrault2021molecular}
Pauline~J Ollitrault, Alexander Miessen, and Ivano Tavernelli.
\newblock Molecular quantum dynamics: A quantum computing perspective.
\newblock {\em Accounts of Chemical Research}, 54(23):4229--4238, 2021.

\bibitem{rossmannek2021quantum}
Max Rossmannek, Panagiotis~Kl Barkoutsos, Pauline~J Ollitrault, and Ivano
  Tavernelli.
\newblock Quantum hf/dft-embedding algorithms for electronic structure
  calculations: Scaling up to complex molecular systems.
\newblock {\em The Journal of Chemical Physics}, 154(11):114105, 2021.

\bibitem{Panos_Alchemical_2021}
Panagiotis~Kl. Barkoutsos, Fotios Gkritsis, Pauline~J. Ollitrault, Igor~O.
  Sokolov, Stefan Woerner, and Ivano Tavernelli.
\newblock Quantum algorithm for alchemical optimization in material design.
\newblock {\em Chem. Sci.}, 12:4345--4352, 2021.

\bibitem{robert2021resource}
Anton Robert, Panagiotis~Kl Barkoutsos, Stefan Woerner, and Ivano Tavernelli.
\newblock Resource-efficient quantum algorithm for protein folding.
\newblock {\em npj Quantum Information}, 7(1):38, 2021.

\bibitem{mensa2022quantum}
Stefano Mensa, Emre Sahin, Francesco Tacchino, Panagiotis~Kl Barkoutsos, and
  Ivano Tavernelli.
\newblock Quantum machine learning framework for virtual screening in drug
  discovery: a prospective quantum advantage.
\newblock {\em arXiv preprint arXiv:2204.04017}, 2022.

\bibitem{baiardi2022quantum}
Alberto Baiardi, Matthias Christandl, and Markus Reiher.
\newblock Quantum computing for molecular biology.
\newblock {\em arXiv preprint arXiv:2212.12220}, 2022.

\bibitem{maniscalco2022quantum}
Sabrina Maniscalco, Elsi-Mari Borrelli, Daniel Cavalcanti, Caterina Foti, Adam
  Glos, Mark Goldsmith, Stefan Knecht, Keijo Korhonen, Joonas Malmi, Anton
  Nyk{\"a}nen, et~al.
\newblock Quantum network medicine: rethinking medicine with network science
  and quantum algorithms.
\newblock {\em arXiv preprint arXiv:2206.12405}, 2022.

\bibitem{kass2008}
Ivan Kassal, Stephen~P. Jordan, Peter~J. Love, Masoud Mohseni, and Alán
  Aspuru-Guzik.
\newblock Polynomial-time quantum algorithm for the simulation of chemical
  dynamics.
\newblock {\em Proceedings of the National Academy of Sciences},
  105(48):18681--18686, 2008.

\bibitem{veis2016b}
{L.} Veis, {J.} Visnak, {H.} Nishizawa, {H.} Nakai, and {J.} Pittner.
\newblock Quantum chemistry beyond born–oppenheimer approximation on a
  quantum computer: A simulated phase estimation study.
\newblock {\em Int. J. Quantum. Chem.}, 116:1328--1336, 2016.

\bibitem{olli2020}
Pauline~J. Ollitrault, Guglielmo Mazzola, and Ivano Tavernelli.
\newblock Nonadiabatic molecular quantum dynamics with quantum computers.
\newblock {\em Phys. Rev. Lett.}, 125:260511, Dec 2020.

\bibitem{pavo2021}
Fabijan Pavošević and Sharon Hammes-Schiffer.
\newblock Multicomponent unitary coupled cluster and equation-of-motion for
  quantum computation.
\newblock {\em Journal of Chemical Theory and Computation}, 17(6):3252--3258,
  2021.
\newblock PMID: 33945684.

\bibitem{peru2014}
Alberto Peruzzo, Jarrod McClean, Peter Shadbolt, Man-Hong Yung, Xiao-Qi Zhou,
  Peter~J. Love, Alán Auspuru-Guzik, and Jeremy~L. O’Brien.
\newblock A variational eigenvalue solver on a photonic quantum processor.
\newblock {\em Nat. Commun.}, 5:4213, 2014.

\bibitem{moll2018}
Nikolaj Moll, Panagiotis Barkoutsos, Lev~S Bishop, Jerry~M Chow, Andrew Cross,
  Daniel~J Egger, Stefan Filipp, Andreas Fuhrer, Jay~M Gambetta, Marc Ganzhorn,
  Abhinav Kandala, Antonio Mezzacapo, Peter M\"uller, Walter Riess, Gian Salis,
  John Smolin, Ivano Tavernelli, and Kristan Temme.
\newblock Quantum optimization using variational algorithms on near-term
  quantum devices.
\newblock {\em Quantum Sci.Technol.}, 3:030503, 2018.

\bibitem{kand2017}
Abhinav Kandala, Antonio Mezzacapo, Kristan Temme, Maika Takita, Markus Brink,
  {Jerry M.} Chow, and {Jay M.} Gambetta.
\newblock Hardware-efficient variational quantum eigensolver for small
  molecules and quantum magnets.
\newblock {\em Nature}, 249:242, 2017.

\bibitem{seeley2012bravyi}
Jacob~T Seeley, Martin~J Richard, and Peter~J Love.
\newblock The bravyi-kitaev transformation for quantum computation of
  electronic structure.
\newblock {\em The Journal of chemical physics}, 137(22):224109, 2012.

\bibitem{mcardle2020quantum}
Sam McArdle, Suguru Endo, Al{\'a}n Aspuru-Guzik, Simon~C Benjamin, and Xiao
  Yuan.
\newblock Quantum computational chemistry.
\newblock {\em Reviews of Modern Physics}, 92(1):015003, 2020.

\bibitem{preskill2018quantum}
John Preskill.
\newblock Quantum computing in the nisq era and beyond.
\newblock {\em Quantum}, 2:79, 2018.

\bibitem{mcclean2018barren}
Jarrod~R McClean, Sergio Boixo, Vadim~N Smelyanskiy, Ryan Babbush, and Hartmut
  Neven.
\newblock Barren plateaus in quantum neural network training landscapes.
\newblock {\em Nature communications}, 9(1):1--6, 2018.

\bibitem{qiskit}
{Qiskit}.
\newblock Qiskit: An open-source framework for quantum computing, 2021.
\newblock http://www.qiskit.org.

\bibitem{wang2021noise}
Samson Wang, Enrico Fontana, Marco Cerezo, Kunal Sharma, Akira Sone, Lukasz
  Cincio, and Patrick~J Coles.
\newblock Noise-induced barren plateaus in variational quantum algorithms.
\newblock {\em Nature communications}, 12(1):1--11, 2021.

\bibitem{GAMESS}
Giuseppe M.~J. Barca, Colleen Bertoni, Laura Carrington, Dipayan Datta, Nuwan
  De~Silva, J.~Emiliano Deustua, Dmitri~G. Fedorov, Jeffrey~R. Gour,
  Anastasia~O. Gunina, Emilie Guidez, Taylor Harville, Stephan Irle, Joe
  Ivanic, Karol Kowalski, Sarom~S. Leang, Hui Li, Wei Li, Jesse~J. Lutz, Ilias
  Magoulas, Joani Mato, Vladimir Mironov, Hiroya Nakata, Buu~Q. Pham, Piotr
  Piecuch, David Poole, Spencer~R. Pruitt, Alistair~P. Rendell, Luke~B. Roskop,
  Klaus Ruedenberg, Tosaporn Sattasathuchana, Michael~W. Schmidt, Jun Shen,
  Lyudmila Slipchenko, Masha Sosonkina, Vaibhav Sundriyal, Ananta Tiwari,
  Jorge~L. Galvez~Vallejo, Bryce Westheimer, Marta Wloch, Peng Xu, Federico
  Zahariev, and Mark~S. Gordon.
\newblock Recent developments in the general atomic and molecular electronic
  structure system.
\newblock {\em The Journal of Chemical Physics}, 152(15):154102, April 2020.

\bibitem{gommersscipy}
Ralf Gommers, Pauli Virtanen, Evgeni Burovski, Warren Weckesser, Travis~E
  Oliphant, David Cournapeau, Matt Haberland, Tyler Reddy, Pearu Peterson,
  Andrew Nelson, et~al.
\newblock scipy/scipy: Scipy 1.8.0.
\newblock {\em Zenodo}, 2022.

\bibitem{powell1994direct}
Michael~JD Powell.
\newblock A direct search optimization method that models the objective and
  constraint functions by linear interpolation.
\newblock In {\em Advances in optimization and numerical analysis}, pages
  51--67. Springer, 1994.

\bibitem{hest1952}
Magnus~R. Hestenes and Eduard Stiefel.
\newblock Methods of conjugate gradients for solving linear systems.
\newblock {\em Journal of Research of the National Bureau of Standards},
  49(6):409, 1952.

\bibitem{nazareth2009conjugate}
John~L Nazareth.
\newblock Conjugate gradient method.
\newblock {\em Wiley Interdisciplinary Reviews: Computational Statistics},
  1(3):348--353, 2009.

\bibitem{kraft1994slsqp}
Dieter Kraft.
\newblock Algorithm 733: {TOMP–Fortran} modules for optimal control
  calculations.
\newblock {\em ACM Trans. Math. Softw.}, 20(3):262–281, sep 1994.

\bibitem{brav2017}
Sergey Bravyi, Jay~M. Gambetta, Antonio Mezzacapo, and Kristan Temme.
\newblock Tapering off qubits to simulate fermionic hamiltonians.
\newblock {\em arXiv:1701.08213}, 2017.

\bibitem{631G}
R.~Ditchfield, W.~J. Hehre, and J.~A. Pople.
\newblock Self‐consistent molecular‐orbital methods. ix. an extended
  gaussian‐type basis for molecular‐orbital studies of organic molecules.
\newblock {\em The Journal of Chemical Physics}, 54(2):724--728, 1971.

\bibitem{hube1979}
Huber K.P. and Herzberg G.
\newblock {M}olecular {S}pectra and {M}olecular {S}tructure. {IV}. {C}onstants
  of {D}iatomic {M}olecules.
\newblock {National Institute of Standards and Technology}, 1979.

\bibitem{iord2003}
Tzvetelin Iordanov and Sharon Hammes-Schiffer.
\newblock {Vibrational analysis for the nuclear-electronic orbital method}.
\newblock {\em Journal of Chemical Physics}, 118(21):9489--9496, 2003.

\bibitem{hurl1988}
{A.} Hurley.
\newblock The computation of floating functions and their use in force constant
  calculations.
\newblock {\em Journal of Computational Chemistry}, 9(1):75--79, 1988.

\bibitem{tach1999}
Masanori Tachikawa, Kento Taneda, and Kazuhide Mori.
\newblock Simultaneous optimization of gtf exponents and their centers with
  fully variational treatment of hartree–fock molecular orbital calculation.
\newblock {\em International Journal of Quantum Chemistry}, 75(4-5):497--510,
  1999.

\bibitem{naka2001}
Hiromi Nakai, Keitaro Sodeyama, and Minoru Hoshino.
\newblock {N}on-{B}orn–{O}ppenheimer theory for simultaneous determination of
  vibrational and electronic excited states: ab initio {NO+MO}/{CIS} theory.
\newblock {\em Chemical Physics Letters}, 345(1):118--124, 2001.

\bibitem{naka2005}
Hiromi Nakai, Minoru Hoshino, Kaito Miyamoto, and Shiaki Hyodo.
\newblock Elimination of translational and rotational motions in nuclear
  orbital plus molecular orbital theory.
\newblock {\em The Journal of Chemical Physics}, 122(16):164101, 2005.

\bibitem{bror2020}
Kurt~R. Brorsen.
\newblock {Quantifying Multireference Character in Multicomponent Systems with
  Heat-Bath Configuration Interaction}.
\newblock {\em Journal of Chemical Theory and Computation}, 16(4):2379--2388,
  2020.

\bibitem{pavo2020}
Fabijan Pavo{\v{s}}evi{\'{c}}, Tanner Culpitt, and Sharon Hammes-Schiffer.
\newblock {Multicomponent Quantum Chemistry: Integrating Electronic and Nuclear
  Quantum Effects via the Nuclear-Electronic Orbital Method}.
\newblock {\em Chemical Reviews}, 120(9):4222--4253, 2020.

\bibitem{tach2000}
Masanori Tachikawa and Yoshihiro Osamura.
\newblock {Isotope effect of hydrogen and lithium hydride molecules.
  Application of the dynamic extended molecular orbital method and energy
  component analysis}.
\newblock {\em Theoretical Chemistry Accounts}, 104(1):29--39, 2000.

\bibitem{kolo1963}
W.~Ko\l{}os and L.~Wolniewicz.
\newblock Nonadiabatic theory for diatomic molecules and its application to the
  hydrogen molecule.
\newblock {\em Rev. Mod. Phys.}, 35:473--483, Jul 1963.

\bibitem{muol2019}
Andrea Muolo, Edit M{\'{a}}tyus, and Markus Reiher.
\newblock {H$_3^+$ as a five-body problem described with explicitly correlated
  Gaussian basis sets}.
\newblock {\em Journal of Chemical Physics}, 151(15), 2019.

\bibitem{wang2008a}
Yimin Wang, Bastiaan~J. Braams, Joel~M. Bowman, Stuart Carter, and David~P.
  Tew.
\newblock Full-dimensional quantum calculations of ground-state tunneling
  splitting of malonaldehyde using an accurate ab initio potential energy
  surface.
\newblock {\em The Journal of Chemical Physics}, 128(22):224314, 2008.

\bibitem{list2020}
Nanna~H. List, Adrian~L. Dempwolff, Andreas Dreuw, Patrick Norman, and Todd~J.
  Martínez.
\newblock Probing competing relaxation pathways in malonaldehyde with transient
  x-ray absorption spectroscopy.
\newblock {\em Chem. Sci.}, 11:4180--4193, 2020.

\bibitem{baug1981}
Steven~L. Baughcum, Richard~W. Duerst, Walter~F. Rowe, Zuzana Smith, and
  E.~Bright Wilson.
\newblock {Microwave Spectroscopic Study of Malonaldehyde
  (3-Hydroxy-2-propenal). 2. Structure, Dipole Moment, and Tunneling}.
\newblock {\em Journal of the American Chemical Society}, 103(21):6296--6303,
  1981.

\bibitem{flud1977}
Eugene~{M.} Fluder and {Jose R.} {De La Vega}.
\newblock Intramolecular hydrogen tunneling in malonaldehyde.
\newblock {\em J. Am. Chem. Soc.}, 100(17):5265--5267, 1977.

\bibitem{moll1934}
{Chr.} M{\o}ller and {M. S.} Plesset.
\newblock {Note on an approximation treatment for many-electron systems}.
\newblock {\em Physical Review}, 46(7):618--622, 1934.

\bibitem{yang2019}
Yang Yang, Patrick~E. Schneider, Tanner Culpitt, Fabijan Pavo{\v{s}}evi{\'{c}},
  and Sharon Hammes-Schiffer.
\newblock {Molecular Vibrational Frequencies within the Nuclear-Electronic
  Orbital Framework}.
\newblock {\em Journal of Physical Chemistry Letters}, 10(6):1167--1172, 2019.

\bibitem{char2013}
Zachary~B Charles, Miriam Farber, Charles~R Johnson, and Lee Kennedy-Shaffer.
\newblock The relation between the diagonal entries and the eigenvalues of a
  symmetric matrix, based upon the sign pattern of its off-diagonal entries.
\newblock {\em Linear Algebra and its Applications}, 438:1427--1445, 2013.

\bibitem{lege2006}
Legeza \"O., Gebhard F., and Rissler J.
\newblock Entanglement production by independent quantum channels.
\newblock {\em Phys. Rev. B}, 74:195112, 2006.

\bibitem{bogu2012b}
Katharina Boguslawski, Pawel Tecmer, \"Ors Legeza, and Markus Reiher.
\newblock Entanglement measures for single- and multireference correlation
  effects.
\newblock {\em J. Phys. Chem. Lett.}, 3:3129--3135, 2012.

\bibitem{bogu2015}
Katharina Boguslawski and Pawel Tecmer.
\newblock Orbital entanglement in quantum chemistry.
\newblock {\em Int. J. Quantum Chem.}, 115:1289--1295, 2015.

\end{thebibliography}

\end{document}